\documentclass[twocolumn,authoryear,final,11pt,times]{elsarticle}

\usepackage{amssymb}
\usepackage{lineno}
\usepackage{multirow}
\usepackage{longtable}
\usepackage{hhline}
\usepackage{booktabs} 
\usepackage{array}
\usepackage{booktabs} 
\usepackage{listings}
\usepackage{amsmath}
\usepackage{graphicx}
\usepackage{rotating}
\usepackage{tikz}
\usepackage{subcaption}
\usepackage{hyperref}
\usepackage{float}

\journal{Building and Environment}

\begin{document}

\begin{frontmatter}


\title{Make yourself comfortable: Nudging urban heat and noise mitigation with smartwatch-based Just-in-time Adaptive Interventions (JITAI)}

\author[dbe,cis]{Clayton Miller\corref{cor1}}
\author[dbe]{Yun Xuan Chua}
\author[dbe,sec]{Matias Quintana}
\author[doa]{Binyu Lei}
\author[doa,dre]{Filip Biljecki}
\author[dbe,doa,bear]{Mario Frei}

\affiliation[dbe]{organization={Department of the Built Environment, National University of Singapore}, 
            postcode={117566}, 
            country={Singapore}}
\affiliation[cis]{organization={College of Integrative Studies, Singapore Management University}, 
            postcode={179873}, 
            country={Singapore}}
\affiliation[sec]{organization={Future Cities Lab Global Programme, Singapore-ETH Centre},
        postcode={138602}, 
        country={Singapore}}
\affiliation[doa]{organization={Department of Architecture, National University of Singapore}, 
            postcode={117566}, 
            country={Singapore}}
\affiliation[dre]{organization={Department of Real Estate, National University of Singapore}, 
            postcode={119245}, 
            country={Singapore}}
\cortext[cor1]{Corresponding author (cmiller@smu.edu.sg)}
\affiliation[bear]{organization={Berkeley Education Alliance for Research in Singapore (BEARS)}, 
            postcode={138602},
            country={Singapore}}

\begin{abstract}
Humans can play a more active role in improving their comfort in the built environment if given the right information at the right place and time. This paper outlines the use of Just-in-Time Adaptive Interventions (JITAI) implemented in the context of the built environment to provide information that helps humans minimize the impact of heat and noise on their daily lives. This framework is based on the open-source Cozie iOS smartwatch platform. It includes data collection through micro-surveys and intervention messages triggered by environmental, contextual, and personal history conditions. An eight-month deployment of the method was completed in Singapore with 103 participants who submitted more than 12,000 micro-surveys and had more than 3,600 JITAI intervention messages delivered to them. A weekly survey conducted during two deployment phases revealed an overall increase in perceived usefulness ranging from 8-19\% over the first three weeks of data collection. For noise-related interventions, participants showed an overall increase in location changes ranging from 4-11\% and a 2-17\% increase in earphone use to mitigate noise distractions. For thermal comfort-related interventions, participants demonstrated a 3-13\% increase in adjustments to their location or thermostat to feel more comfortable. The analysis found evidence that personality traits (such as conscientiousness), gender, and environmental preferences could be factors in determining the perceived helpfulness of JITAIs and influencing behavior change. These findings underscore the importance of tailoring intervention strategies to individual traits and environmental conditions, setting the stage for future research to refine the delivery, timing, and content of intervention messages.

\end{abstract}



\end{frontmatter}


\section{Introduction}
A key role for the built environment is to provide spaces for work, recreation, socialization, and living that are conducive to human satisfaction \citep{Heydarian2020-is}. Indoor and outdoor environments are designed to optimize thermal, acoustic, visual and other types of comfort, and active climate control and lighting systems are installed to reinforce comfort \citep{Salim2020-ec}. The human role in these systems is often relegated to that of a passive actor for whom one-size-fits-all comfort conditions are delivered in the right place and time \citep{Day2020-tq}. This philosophy underpins much of the thermal comfort research for buildings and cities and strongly emphasizes delivering comfort through sophisticated mechanical systems, the most common being heating and cooling. The fallacy of this approach is that there is an expense and carbon impact to installing and operating these systems, and \emph{delivering comfort} is imprecise due to differences in occupant preference and the limited resolution of comfort zones \citep{Xu2023-mi}. This strategy results in a significant amount of dissatisfaction and poor comfort prediction \citep{Graham2021-en}, and these systems are designed based on thermal comfort prediction models that are only accurate 33\% of the time \citep{Cheung2019-fs}. These models do not account for thermal sensitivity \citep{Rupp2022-lq} and thermal expectation \citep{Schweiker2020-pw}, in which changes in temperature affect occupants in different ways. In addition, thermal comfort and its impact on productivity are more complex than previously thought \citep{Castaldo2018-ih}.

Thermal comfort is only one component of the occupant experience, as noise and distraction are equally or even more critical. A study of the survey results of 600 offices shows that 81\% of the respondents expressed dissatisfaction with at least one aspect of their workspace, with sound privacy and noise-based distraction being among the most common complaints \citep{Parkinson2023-yk}. Mitigation measures are being developed to reduce the impact of noise, but the limited ability to create a diversity of types of office space reduces the ability to reduce noise problems in practice \citep{Kim2013-gs, Appel-Meulenbroek2020-bp}. It is not just the office where people are having these issues; noise and distraction are major considerations in schools \citep{Zipf2020-br}, outdoors \citep{Alias2019-le}, and at home \citep{Bergefurt2023-fm}.

The global COVID-19 pandemic was a catalyst for changes in how people spend time during a typical day \citep{Kumar2021-zm}. For example, it is not a given that office workers commute first thing in the morning, spend eight hours at their individually assigned desks, and then go home in the late afternoon. These worker types often have the choice to work in their homes, in remote, decentralized hub-style coworking offices, or in public spaces such as coffee shops, or even outdoors. There is evidence that this paradigm shift results in increased productivity in some cases \citep{Angelici2023-rw}. This increase in flexibility in the daily lives of people in the built environment sets the foundation for more choices about where to work, play, or spend time with others that can be influenced with information that empowers \emph{choosing of comfort} rather than delivery \citep{Sood2020-vz}.

\subsection{Using JITAI for noise and heat issues at the urban-scale}
Providing people with information at the right place and time to make the best decisions is a technique known as just-in-time adaptive interventions (JITAI) that has significant momentum in the fields of reducing sedentary behavior \citep{Muller2017-ao, Saponaro2021-rx}, promoting physical activity \citep{Hardeman2019-gs, Venema2018-ft} and supporting health behaviors \citep{Nahum-Shani2018-hw}. A significant recent study reviews the use of these types of intervention in the management of Type 2 diabetes \citep{Keller2022-uo}. There are some examples of nudging behavior in the built environment that are similar to these types of studies, but do not use JITAI specifically as a strategy \citep{Soomro2021-ne,Li2021-ej}.

This paper outlines the deployment of JITAI in indoor and outdoor urban settings using smartwatches and the open-source Cozie Apple platform \citep{Tartarini2023-iu}. The high-level conceptual framework was described in previous works \citep{Miller2022-dy, Miller2023-wi}, and this paper provides an in-depth analysis of behavior interventions. This work seeks to address whether JITAI, as a method, is tolerable in the context of the objectives of the built environment and whether providing intervention information in this way can affect occupant behavior in indoor environments. This work relates to the convergence of spatial parameters, sensor and wearable data, and occupant information \citep{Miller2021-bm, Abboushi2022-ot}, which has been used in previous studies indoors \citep{Pollard2022-gl}. It draws inspiration from the use of smartwatches in the urban context, which have been used to characterize urban contexts \citep{Mosteiro-Romero2024-eb}, university campuses \citep{Mosteiro-Romero2023-us,Tekler2023-ri}, and to capture exposure to noise \citep{Fischer2022-lv} and types of interactions \citep{Maisha2023-dt, Sonta2020-pd} in various settings.

\section{Methodology}

The focus of this study was to implement a smartwatch-based data collection process that facilitates intervention messages to potentially change people's behavior and improve their situation. This section outlines the JITAI framework and the details of implementation in the deployment of 103 participants in Singapore. The group of participants was 61\% female and 39\% male, with a majority between 18 and 30 years of age (72\%), followed by smaller groups in the range of 31 to 70 years.

\subsection{Smartwatch micro-survey platform}
\begin{figure*}
    \centering
    \includegraphics[width=\textwidth]{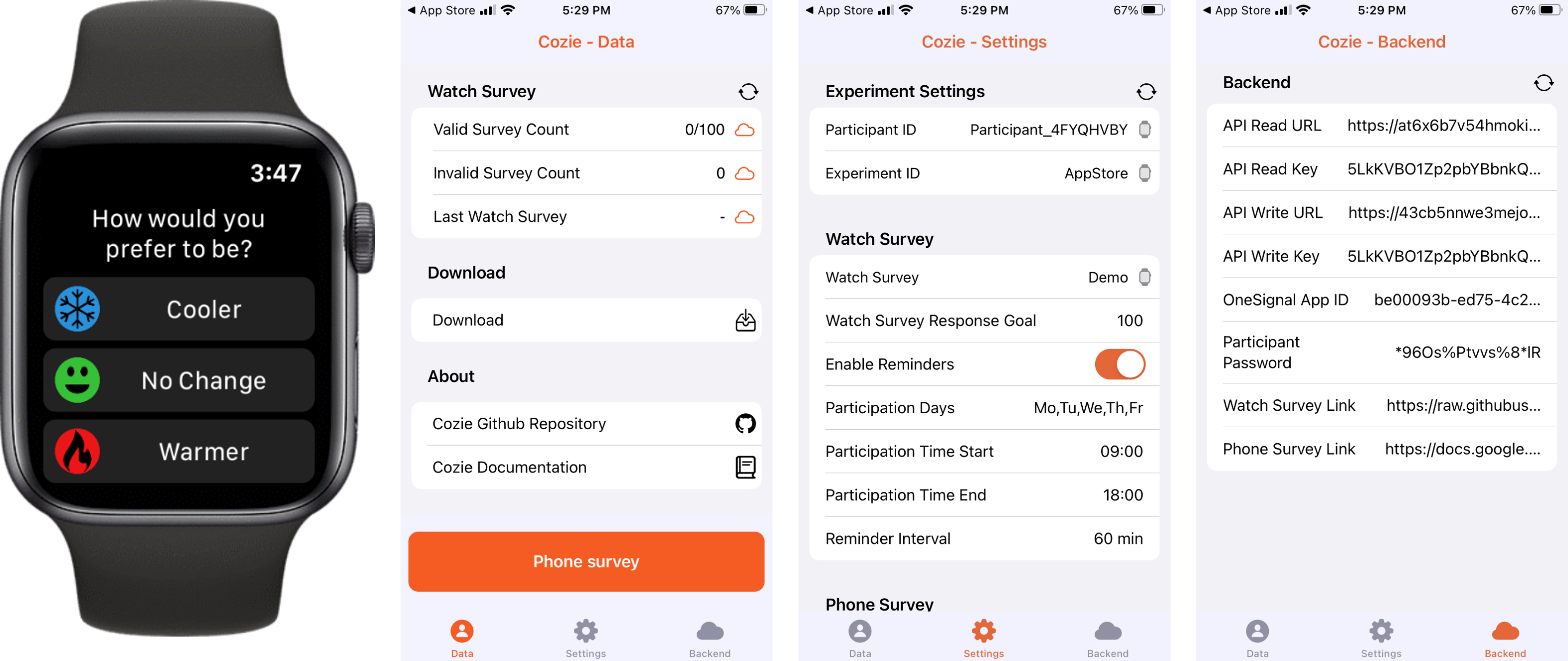}
    \caption{Overview of smartwatch and smartphone interfaces and experimental setup options. More details and documentation of the Cozie Apple platform can be found at: \url{https://cozie-apple.app}}
    \label{fig:cozie_Screenshots}
\end{figure*}

This work uses a smartwatch-based data collection platform\footnote{\url{https://www.cozie-apple.app/}} that uses an iPhone and an Apple Watch to collect micro-survey feedback and physiological data. The iPhone app allows you to change the settings of the app, transfer platform data to the online database, receive push notifications, and display a weekly survey. Screenshots of the application on the Apple Watch and iPhone are shown in Figure \ref{fig:cozie_Screenshots}. This platform is the latest iteration of an open source project that started with a Fitbit version in 2019 \citep{Jayathissa2019-kg, Quintana2021-ka}.

\begin{figure*}
    \centering
    \includegraphics[width=\textwidth]{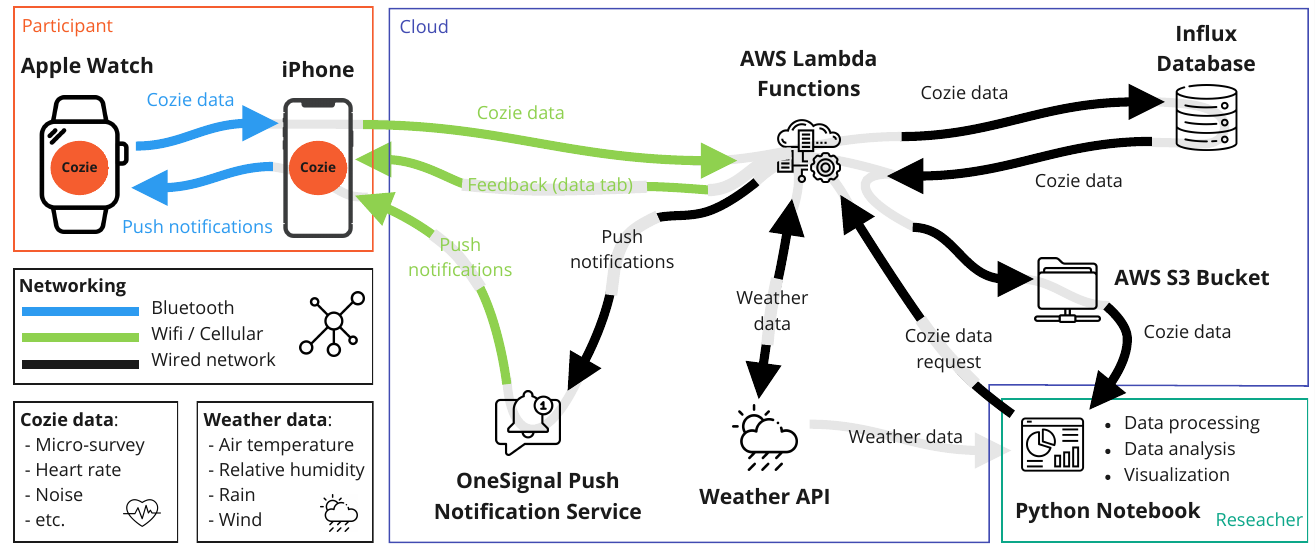}
    \caption{Framework for the data collection, processing, and intervention message delivery.}
    \label{fig:data_flow}
\end{figure*}

The data from this platform are acquired on the Apple Watch (physiological data, watch survey responses) and sent to the iPhone via Bluetooth, where it is relayed to an Amazon Web Services (AWS) Lambda function via Wi-Fi or cellular network. The AWS Lambda function inserts the data into the Influx time-series database. This data transfer occurs within a few seconds of responding to a watch survey. Figure \ref{fig:data_flow} outlines the flow of data throughout the platform.
Cozie data can be retrieved using a Python notebook, where another AWS Lambda function is called via an HTTPS request. The Lambda function retrieves the data from the database and writes them to a CSV file that is saved to an AWS S3 bucket, from which it can then be downloaded to the Python notebook. JITAI messages and other push notifications are scheduled in a Lambda function. This lambda function retrieves Cozie data from the database, weather data from the weather API \footnote{\url{https://beta.data.gov.sg/}}, and then decides which push notification to send. Cozie uses OneSignal as a push notification service provider. Sending a push notification is as simple as making an API call via HTTPS post request. The push notification is then sent from OneSignal to the iPhone and then relayed to the Apple Watch via Bluetooth.

\subsection{Physiological and environmental data acquisition}
The foundation of the decision of whether to serve an intervention message is the data collected from a person from various sources. Within the Cozie Apple framework, there are three categories of data that are collected:
\begin{enumerate}
    \item Micro-survey responses, including geo-location
    \item Physiological and activity data, e.g., heart rate, noise level, step count, etc.
    \item Weather data, for example, outdoor air temperature, rainfall, etc.
\end{enumerate}

\begin{figure*}
    \centering
    \includegraphics[width=\textwidth]{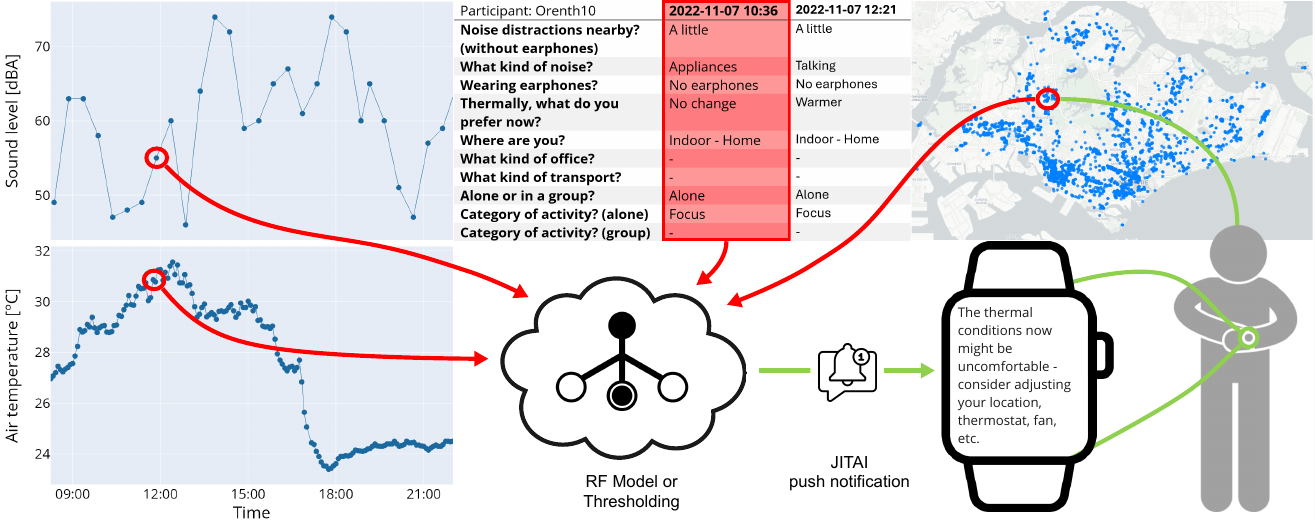}
    \caption{Example of the data fusion from a single participant towards the prompting of the JITAI message. Basemap: (c) OpenStreetMap contributors.} 
    \label{fig:cozie_data}
\end{figure*}

Figure \ref{fig:cozie_data} shows a sample of one day's worth of data.
The participants provide micro-survey responses using the Cozie app on the Apple Watch. The micro-survey response acquires latitude, longitude, and the time of location acquisition. The beginning and end times of the micro-survey are also recorded. The micro-survey responses are sent to the online database immediately after completing the survey.
The Cozie app can also acquire physiological and activity data such as sound level, heart rate, resting heart rate, step count, walking distance, standing time, and oxygen saturation. The Cozie app uses the HealthKit\footnote{\url{https://developer.apple.com/documentation/healthkit}} API provided by Apple to access the data. 
The sampling interval of the physiological and activity data depends on the type of data and can vary over time. Discrete values of equivalent continuous sound pressure level data\footnote{\url{https://developer.apple.com/documentation/healthkit/hkquantitytypeidentifier/3081271-environmentalaudioexposure}} are available at 30-minute intervals, heart rate data at approximately 3-minute intervals, and blood oxygen saturation at approximately 15 minutes. Step count, walking distance, and stand time are sampled at irregular intervals depending on user activity.
The physiological and activity data are sent to the online database in the background, and when the Cozie iPhone app is opened. Background tasks are scheduled by the operating system and depend on system resources and conditions, e.g., battery charge state.
In addition to smartwatch data, air temperature was acquired from a public API in Singapore. All weather data were available at a 1-minute sampling interval. 

\subsection{Experimental deployment}
Participants for this study were recruited through the National University of Singapore (NUS) Student Work Scheme (NSWS) website and word of mouth, with recruitment details primarily shared by NUS students and staff. The participant pool consisted primarily of NUS students, along with a smaller proportion of adults outside the university community. The majority of the participants were 18 to 30, accounting for slightly more than 73\% of the total, and the gender distribution was 39\% male and 61\% female. Ethical approval for the study was obtained from the Institutional Review Board (IRB) of the National University of Singapore.

The participants were equipped with an Apple Watch and the Cozie app was installed on both their iPhones and watches. Participants were required to have an iPhone in order to participate; those who did not have a compatible device were loaned an iPhone SE (2nd generation) for the duration of the study. The majority of participants also owned their own Apple Watch; for those who did not, a loan unit was provided. Participants were asked to wear the watch during working hours from 9\,AM to 7\,PM on weekdays for approximately four weeks. Although this time frame was specified for consistency, participants were not restricted from wearing the watch outside of these hours, and some micro-survey responses may have been submitted beyond the designated window. Each participant was required to complete at least 100 micro-surveys and a weekly survey in the Cozie app, as well as attend an in-person exit interview. These follow-ups were designed to assess the efficacy of the JITAI messages sent to participants during the study period.

\begin{figure}
    \centering
    \includegraphics[clip, trim=0cm 21cm 20.5cm 0cm, width=1\columnwidth]{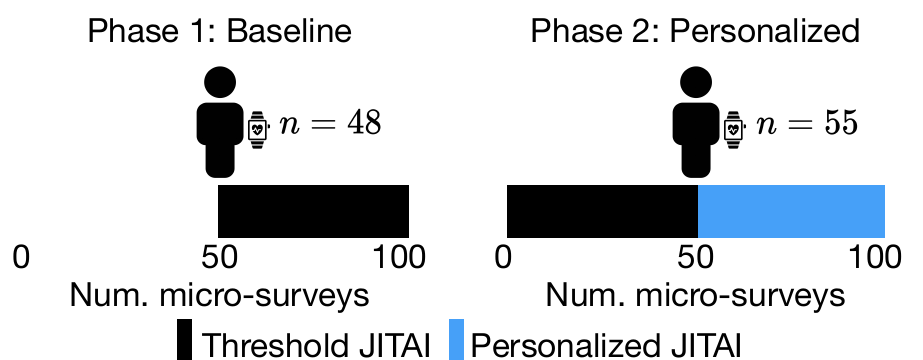}
    \caption{
    Deployment phases: Phases 1 and 2 included 48 and 55 unique participants, respectively.
    Both phases started sending threshold JITAI messages from the beginning, but Phase 2 changed its logic mechanism to personalized JITAI messages after the participant had submitted 50 micro-surveys.}
    \label{fig:phases}
\end{figure}

The study was carried out in two phases (Figure \ref{fig:phases}). 
The first phase involved 48 participants, and its primary goal was to serve as the first attempt at using rule-based logic.
In the second phase, 55 participants were enrolled. 
Similarly to the first phase, rule-based JITAI messages were sent at the beginning of the data collection.
However, in Phase~2, personalized JITAI messages were triggered by the participant completing 50 micro-surveys, which was halfway through the study.
Unlike Phase 1, this new set of messages was predictions based on the first 50 micro-surveys.

\subsection{Overview of JITAI framework}

\begin{figure*}[t]
    \centering
    \includegraphics[width=\textwidth]{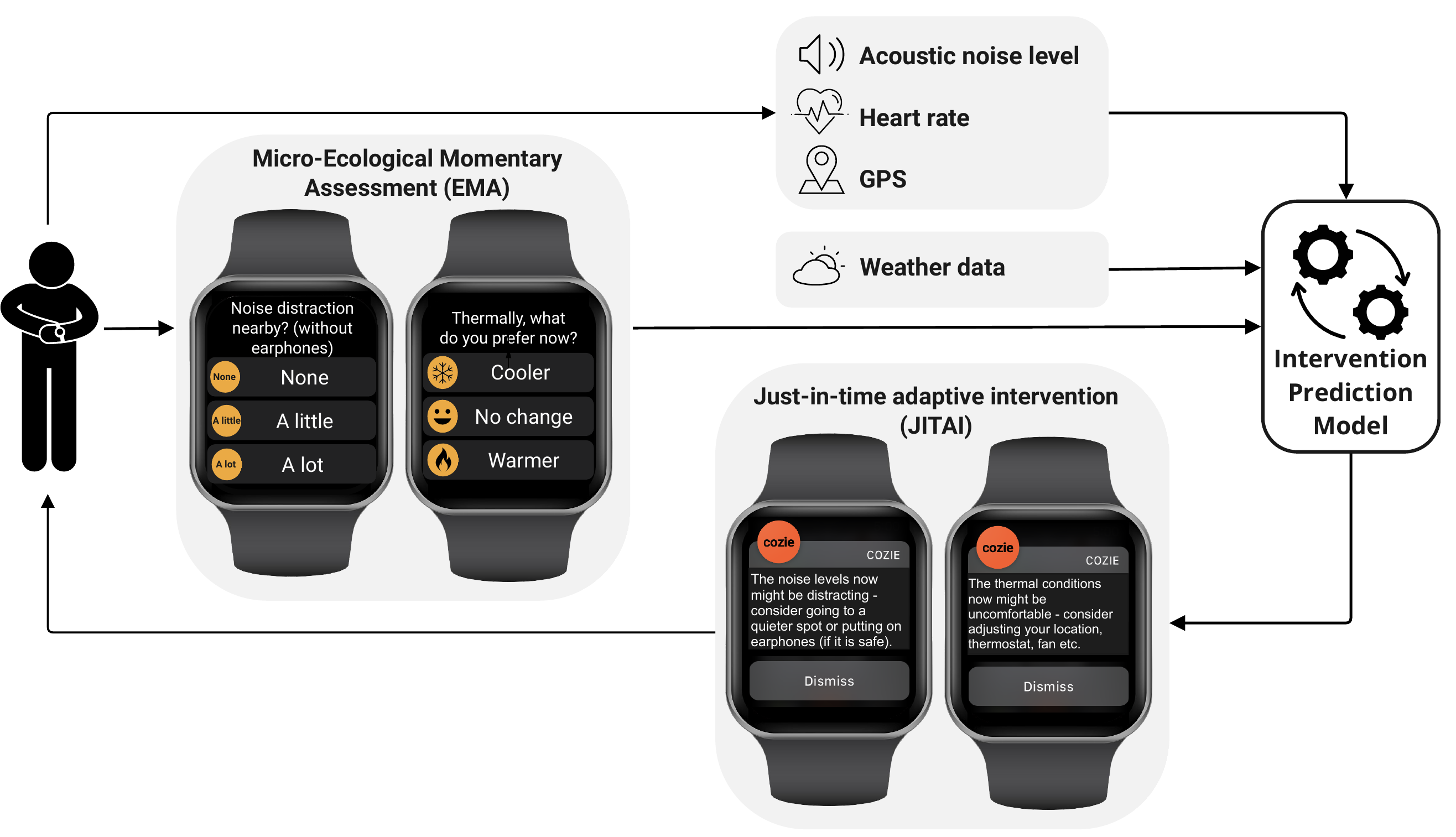}
    \caption{Schematic overview of the smartwatch-based Just-in-Time Adaptive Intervention (JITAI) framework that combines micro-survey responses, physiological data, and environmental inputs to trigger personalized intervention messages using either rule-based or predictive logic.}
    \label{fig:jitai-overview}
\end{figure*}

The smartwatch-based JITAI framework integrates passive sensing (e.g., noise levels, weather, heart rate, GPS) with real-time micro-surveys (e.g., thermal preference, noise distraction) collected via the Cozie app. When participants reported noise distraction, a follow-up prompt asked them to select the dominant sound source (e.g., traffic, talking, weather, etc.). Figure~\ref{fig:jitai-overview} presents the overall framework, illustrating how the physiological, environmental and self-reported data streams are combined to trigger personalized intervention messages. The logic used to determine when these interventions are delivered is described in the following section.

\subsection{JITAI mechanisms}
\begin{figure}
    \centering
    \includegraphics[clip, trim=0cm 21.5cm 20cm 0cm, width=\linewidth]{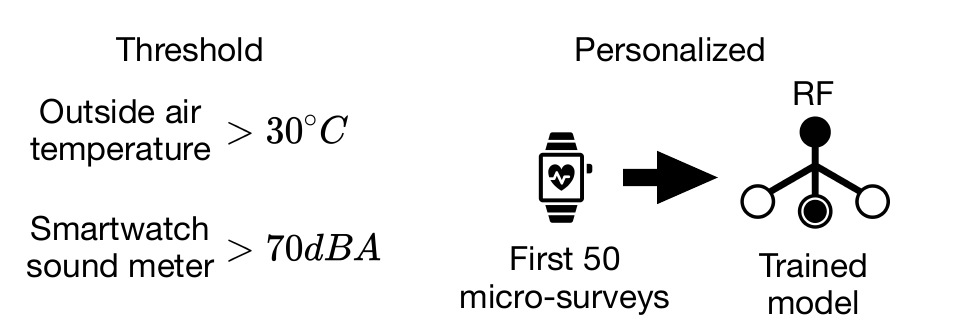}
    \caption{Overview of the JITAI mechanisms for thermal and noise level preference during both deployment phases.
    Threshold-based JITAI messages are based on external conditions, e.g., outside temperature is above 30$^\circ$C and nearby noise is above 70 dBA for thermal and noise level preference, respectively.
    Personalized prediction JITAI messages are based on a prediction model trained on each participant's first 50 micro-survey responses.
    The prediction model used is Random Forest (RF), which predicts the probability that a JITAI message is sent at a given hour.
    }
    \label{fig:jitai-mechanisms}
\end{figure}

While both deployment phases sent JITAI messages throughout the study period, two mechanisms are behind the triggering of the message.
Figure \ref{fig:jitai-mechanisms} shows an overview of the two types of JITAI messages and how they are triggered.
On the one hand, a threshold-based JITAI is triggered when sensor readings exceed a specific threshold.
Thermal preference JITAI messages are sent when the outside air temperature is above 30$^\circ$C.
This threshold is chosen because it is the average outdoor air temperature in Singapore, reflected in the sampled local weather data.
The closest weather station is found based on the GPS location of the smartwatch, and the weather API\footnote{\url{https://beta.data.gov.sg/}} is queried every 5 minutes.
In contrast, the intervention message related to noise is triggered when the smartwatch sound meter readings exceed 70 dBA (left side of Figure \ref{fig:jitai-mechanisms}).
This noise level threshold is chosen because it is in the middle of the decibel range and it is above the upper limit of a normal conversation, that is, 60-70dBA\footnote{\url{https://www.nidcd.nih.gov/health/noise-induced-hearing-loss}}.
Both messages are sent a maximum of four times a day between 9 am and 7 pm on weekdays.
In cases where more than four exceedances occur and/or other push notification reminders are sent, only the first four JITAI messages and reminders are sent, and all remaining triggers are ignored.

The JITAI strategy for personalized prediction is data-driven.
The first 50 micro-surveys were used to train a classification model for each participant's thermal and noise level preference, resulting in two models per participant.
These models used four features: the cumulative distributions of each label's (temperature and noise) class value and its respective hour of the day.
The chosen model is Random Forest (RF), trained with a 3-fold cross-validation, based on its wide usage in personalized comfort prediction~\citep{Quintana2021-ka, Quintana.2023}.
The feature set was kept simple based on the limited amount of data, i.e. 50 training data points, and to have the model solely based on individual preferences.
In other words, for thermal and noise level preference labels, a model will predict the probability of a participant's preference based on their historical trend (i.e., the values of the first 50 micro-survey responses) and hour of the day.

\begin{figure}
    \centering
    \includegraphics[clip, trim=0cm 21cm 17cm 0cm, width=\linewidth]{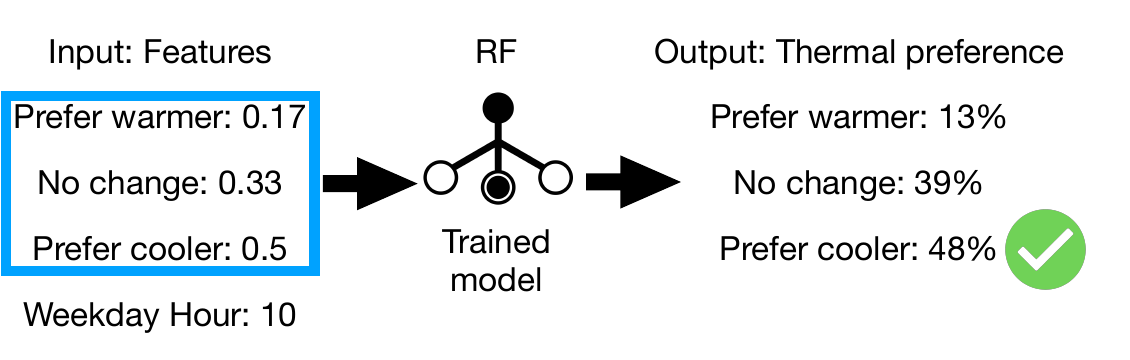}
    \caption{
    Example of a personalized RF model predicting the probabilities of thermal preference using the preference distribution (blue square) and hour of the day as input (10 am).
    The class \emph{Prefer cooler} has the highest probability and thus, a JITAI message will be sent at 10 am to this participant.
    }
    \label{fig:jitai-prediction}
\end{figure}

Figure \ref{fig:jitai-prediction} shows an example of an RF model that predicts the probabilities of thermal preference of a participant based on its trend of historical preference and time of day.
For each remaining day of the Phase 2 deployment, after the first 50 micro-surveys are collected, a participant's personalized RF evaluates all hours between 9 AM and 7 PM as input.
This results in 11 predicted comfort label probabilities, one for each hour.
From these, the four main labels, different from \emph{No Change}, are prioritized to be sent.
The same rationale and workflow are used for noise level preference.

\section{Results}

The deployment of the outlined method occurred from October 2022 until May 2023 and included 103 participants. This section outlines the results of the weekly survey that targeted a subjective evaluation by participants on their perceived effectiveness of JITAI messages, whether their behavior was influenced, and whether there was a sense of annoyance due to the messages sent. The results indicate the personal characteristics that have a potentially significant effect on the tendency of a participant to find the method useful or not. In addition, a spatial analysis emphasizes the approximate influence of where the participants were when the JITAI messages were sent.

\subsection{Perception and tolerance of JITAI}

\begin{figure*}
    \centering
    \includegraphics[width=\textwidth]{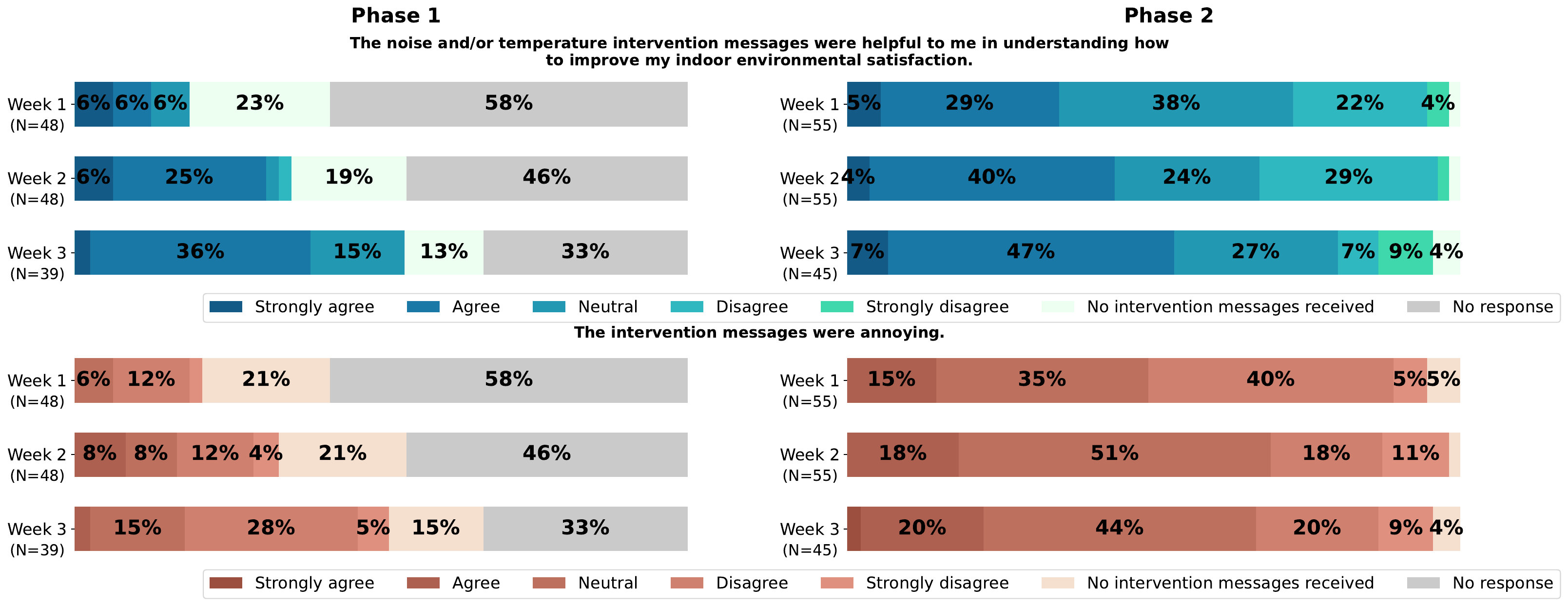}
    \caption{Weekly survey results from Phase 1 (left) and Phase 2 (right), showing the response from Weeks 1-3.}
    \label{fig:orenth_first-weeklysurvey1}
\end{figure*}

The first analysis outlined is the results of the weekly user survey that asked participants about their perceptions of the JITAI method in this context. The survey results from both Phase 1 and Phase 2 reveal a compelling trend in the perceived helpfulness of the intervention messages over a period of three weeks. As depicted in Figure \ref {fig:orenth_first-weeklysurvey1}, Phase 1 begins with 12\% of participants (\emph{Agree} and \emph{Strongly agree} responses combined) acknowledging the usefulness of the intervention messages in the first week. This percentage increased to 31\% in the second week and increased further to 39\% in the third week. Phase 2 shows a similar progression, with an initial agreement level of 34\%, which increased to 44\% in the second week and 54\% in the third week. This upward trend suggests that participants have increasingly recognized the relevance and usefulness of the intervention messages received over time.

Regarding participants who find the intervention messages annoying, the survey results from both Phase 1 and Phase 2 show varying trends. Despite recognizing the usefulness of the messages, there is evidence of fluctuating levels of annoyance reported by the participants. In Phase 1, no participants responded with \emph{Strongly Agree} or \emph{Agree} because they found the messages annoying in Week 1. In Week 2, 8\% of the participants responded with \emph{Agree}, and in Week 3, this percentage dropped to 3\%, indicating a decreasing level of annoyance over time. In contrast, Phase 2 shows a more consistent upward trend. In the first two weeks, 15\% and 18\% of the participants responded, respectively, with \emph{Agree}. By Week 3, the percentage increased significantly to 22\% (\emph{Strongly agree} and \emph{Agree} responses combined). This suggests that while messages are perceived as useful, their repetitive or intrusive nature could contribute to increasing annoyance over time.

\begin{figure*}
    \centering
    \includegraphics[width=\textwidth]{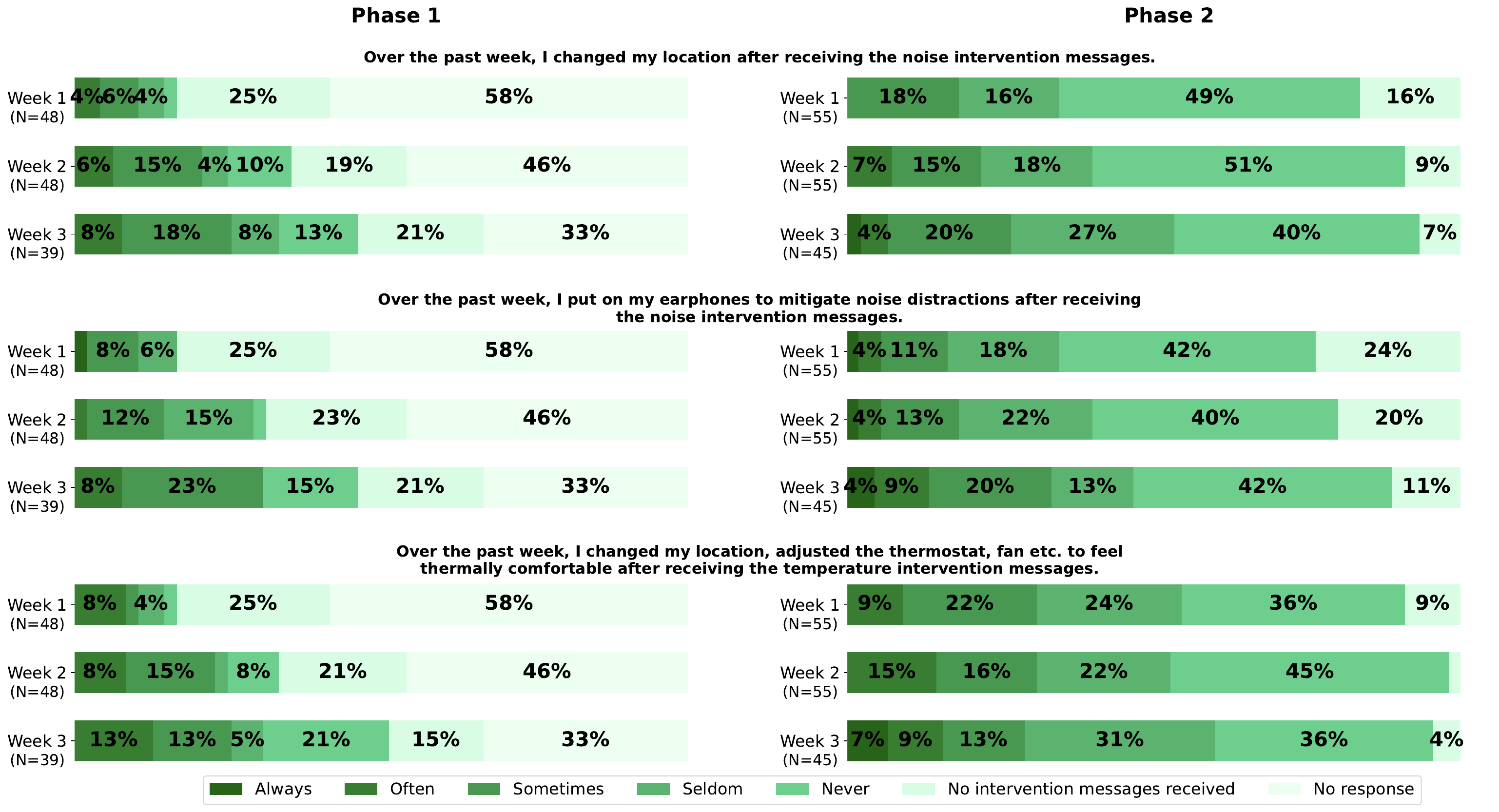}
    \caption{Weekly survey results from Phase 1 (left) and Phase 2 (right), showing the Weeks 1-3 of response for the behavioral change questions.}
    \label{fig:orenth_first-weeklysurvey2}
\end{figure*}

The results shown in Figure \ref{fig:orenth_first-weeklysurvey2} provide information on how individuals responded behaviorally to the noise and temperature intervention messages during the study. Regarding noise mitigation, in Phase 1, there was a slight increase in the number of participants who changed their location due to noise interventions. Among the respondents, 10\% of the participants (\emph{Often} and \emph{Sometimes} combined responses) reported changing their location in the first week. This proportion increased to 21\% in the second week and 26\% in the third week. In Phase 2, there was a gradual increase in participants who reported changing their location due to noise interventions. In the first week, 18\% of the participants reported changing their location (\emph{Sometimes}). By the second week, this increased to 22\% (\emph{Often} and \emph{Sometimes} combined responses). In the third week, the percentage rose further to 26\% (\emph{Always}, \emph{Often} and \emph{Sometimes} combined responses). This trend might suggest that as the study progressed, participants became increasingly likely to change their behavior in response to noise intervention messages.

The usage of earphones showed an upward trend. In the first week, 10\% of the participants reported using earphones (\emph{Always} and \emph{Sometimes} combined responses). This increased to 14\% (\emph{Often} and \emph{Sometimes} combined responses) in the second week and 31\% in the third week. Phase 2 showed a similar upward trend, with 17\% of the participants (\emph{Always}, \emph{Often} and \emph{Sometimes} combined responses) using headphones to mitigate noise distractions in the first week. This increased to 19\% in the second week and further to 33\% in the third week. These results suggest that the intervention messages were effective in encouraging participants to adopt earphone use as a noise mitigation strategy. The substantial and consistent increase in both phases highlights the potential of such interventions to promote positive behavioral changes over time.

Regarding the effectiveness of the thermal comfort intervention messages, the responses in Phase 1 showed that 10\% of the participants (often and sometimes combined) adjusted their behavior in the first week, such as changing locations or modifying thermostat settings. This percentage increased significantly to 23\% in the second week and remained steady at 26\% in the third week (\emph{Always}, \emph{Often} and \emph{Sometimes} combined responses). In Phase 2, 31\% of participants (\emph{Often} and \emph{Sometimes} combined responses) took action to achieve thermal comfort in the first week. This percentage remained at 31\% in the second week and decreased slightly to 29\% in the third week (7\% \emph{Always}, 9\% \emph{Often}, and 13\% \emph{Sometimes}). These findings suggest an adaptive behavioral response among participants. Although Phase 1 shows a steady increase in actions taken, Phase 2 exhibits stabilization after an initially high percentage, potentially reflecting a consistent engagement with the thermal comfort strategies suggested in the intervention messages.

The observed differences in behavioral responses suggest that participants may have found it easier or more habitual to respond to noise-related discomfort than to thermal discomfort. Acoustic interventions showed a consistent increase in behavioral adoption (e.g., location change and earphone use), while thermal intervention responses remained relatively stable after the first week. As this was a field study, the variation in environmental conditions (e.g., noise levels and outdoor temperatures) was not controlled throughout the week. Consequently, it remains unclear whether these behavioral trends were primarily driven by repeated exposure to interventions or by fluctuations in ambient conditions over time.

\subsection{Personal attributes analysis}
\begin{figure*}
    \centering
    \includegraphics[width=\textwidth]{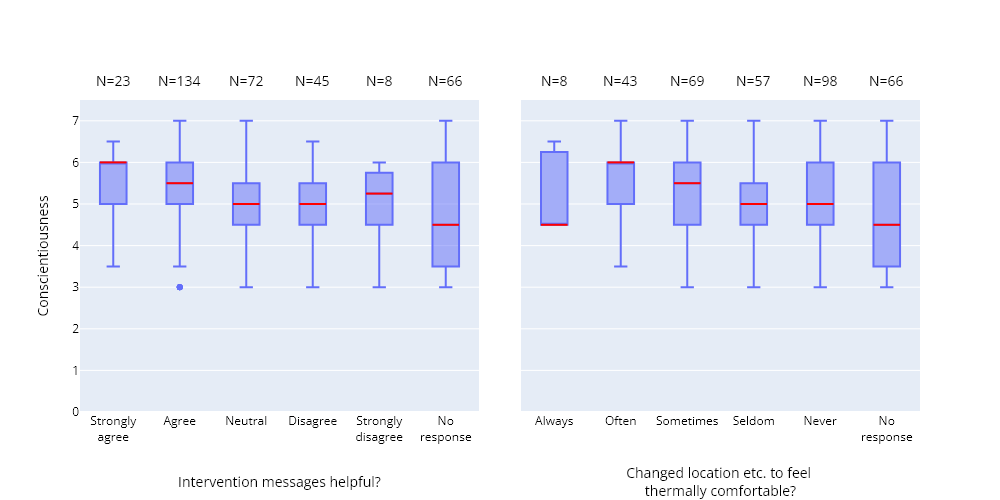}
    \caption{Conscientiousness scores across levels of agreement with perceived helpfulness of JITAI messages (left) and frequency of changing location for thermal comfort (right), with results from both Phase 1 and Phase 2. Participants with higher conscientiousness scores generally show stronger agreement with the helpfulness of JITAI messages. Those who frequently adjusted their location or surroundings to improve thermal comfort also tend to have higher conscientiousness, though sample sizes vary across categories. Responses exclude those who did not receive JITAI messages.}
    \label{fig:Conscientiousness_score_boxplot}
\end{figure*}

To explore the potential impacts of JITAI on the participants, our study segmented the respondents according to their perceptions of the usefulness of the interventions. This analysis aimed to identify attributes within the onboarding survey data that might be associated with variations in participants' responses to the intervention messages. An ordinal logistic regression analysis and a chi-square test were performed to examine how participant demographics and tendencies relate to perceived effectiveness of the intervention messages. The full analyses are available in the open dataset for further exploration. For example, people who scored higher on conscientiousness tended to find the JITAI messages more helpful, as shown in the boxplots to the left of Figure \ref{fig:Conscientiousness_score_boxplot}. Participants who expressed a stronger agreement with the helpfulness of the messages had, on average, higher conscientiousness scores, indicating that personality traits may influence how individuals respond and engage with recommended behaviors.

The plot on the right in Figure \ref{fig:Conscientiousness_score_boxplot} further illustrates this trend, indicating that participants with higher conscientiousness scores may be more inclined to adjust their location or surroundings to improve thermal comfort. Those who reported making these adjustments frequently (for example, \emph{Always} or \emph{Often}) generally had higher conscientiousness scores, suggesting a correlation between personality traits and proactive behavior changes. However, sample sizes varied between categories, which may influence the observed associations.

\begin{figure*}
    \centering
    \includegraphics[width=\textwidth]{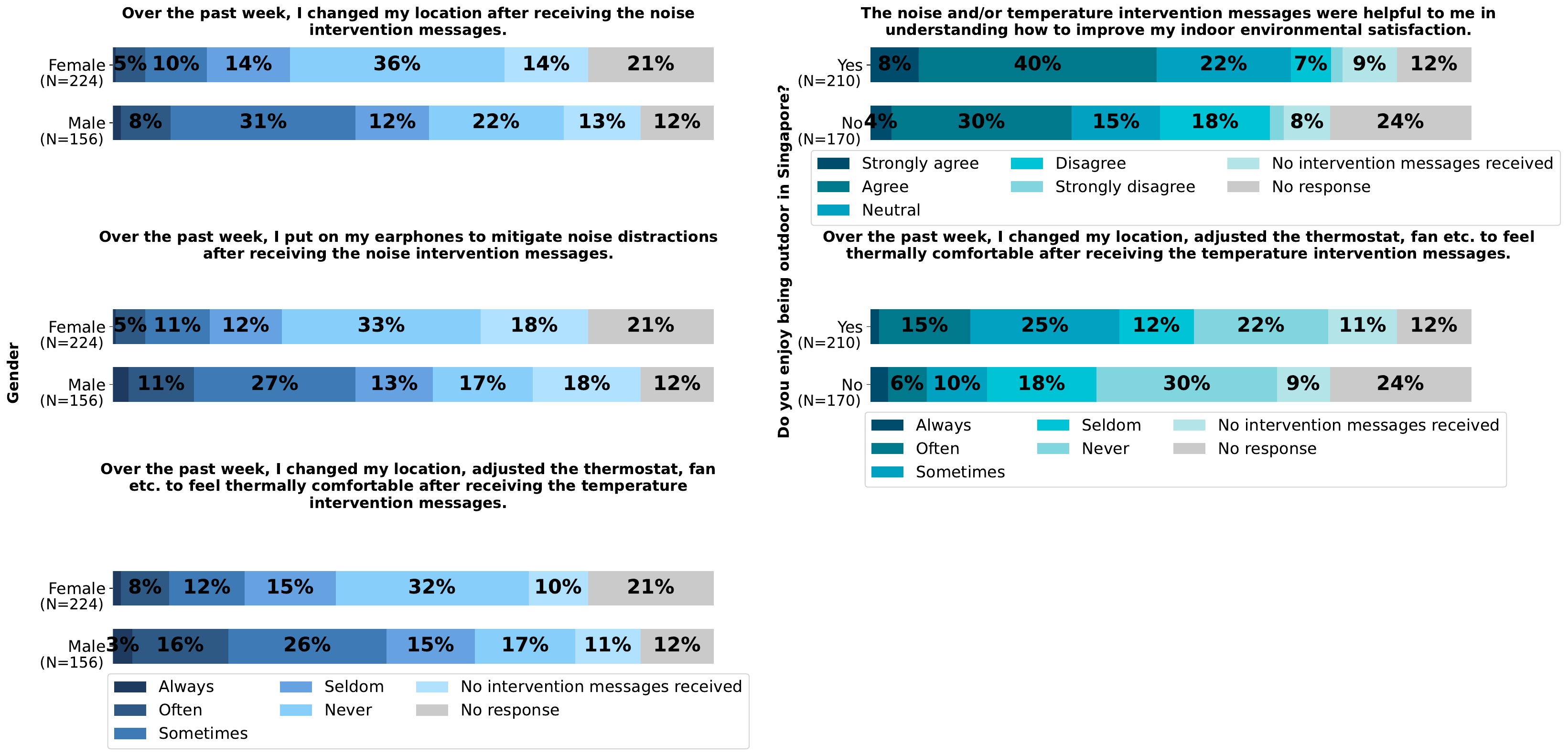}
    \caption{Gender-based responses to location changes and earphone use after noise JITAI messages, and actions after thermal JITAI messages (left); outdoor preferences and their influence on responses to thermal adjustments (right), with results from both Phase 1 and Phase 2. Males reported a higher frequency of location changes and earphone use following noise JITAI messages compared to females. Participants who enjoy being outdoors found thermal JITAI messages more helpful and were less likely to make environmental adjustments. Responses exclude those who did not receive JITAI messages.}
    \label{fig:gender_enjoy_outdoor_charts}
\end{figure*}

Furthermore, gender and environmental preferences appeared to influence the responses of participants to the JITAI messages. As shown in Figure \ref{fig:gender_enjoy_outdoor_charts}, male participants generally reported higher percentages of \emph{Always} and \emph{Often} in multiple behaviors, such as changing location, using headphones to mitigate noise distractions, and adjusting location or thermostat for thermal comfort. Participants who reported enjoying outdoor environments in Singapore appeared to be more likely to adjust their location or thermostat for thermal comfort compared to those who did not enjoy outdoor settings. The stacked bar charts highlight these behavioral differences based on both gender and environmental preferences, suggesting that these factors may have influenced participants' responses to the noise and temperature JITAI messages.

\subsection{Spatial analysis}\label{sec:spatial}
\begin{figure*}
    \centering
    \includegraphics[width=\textwidth]{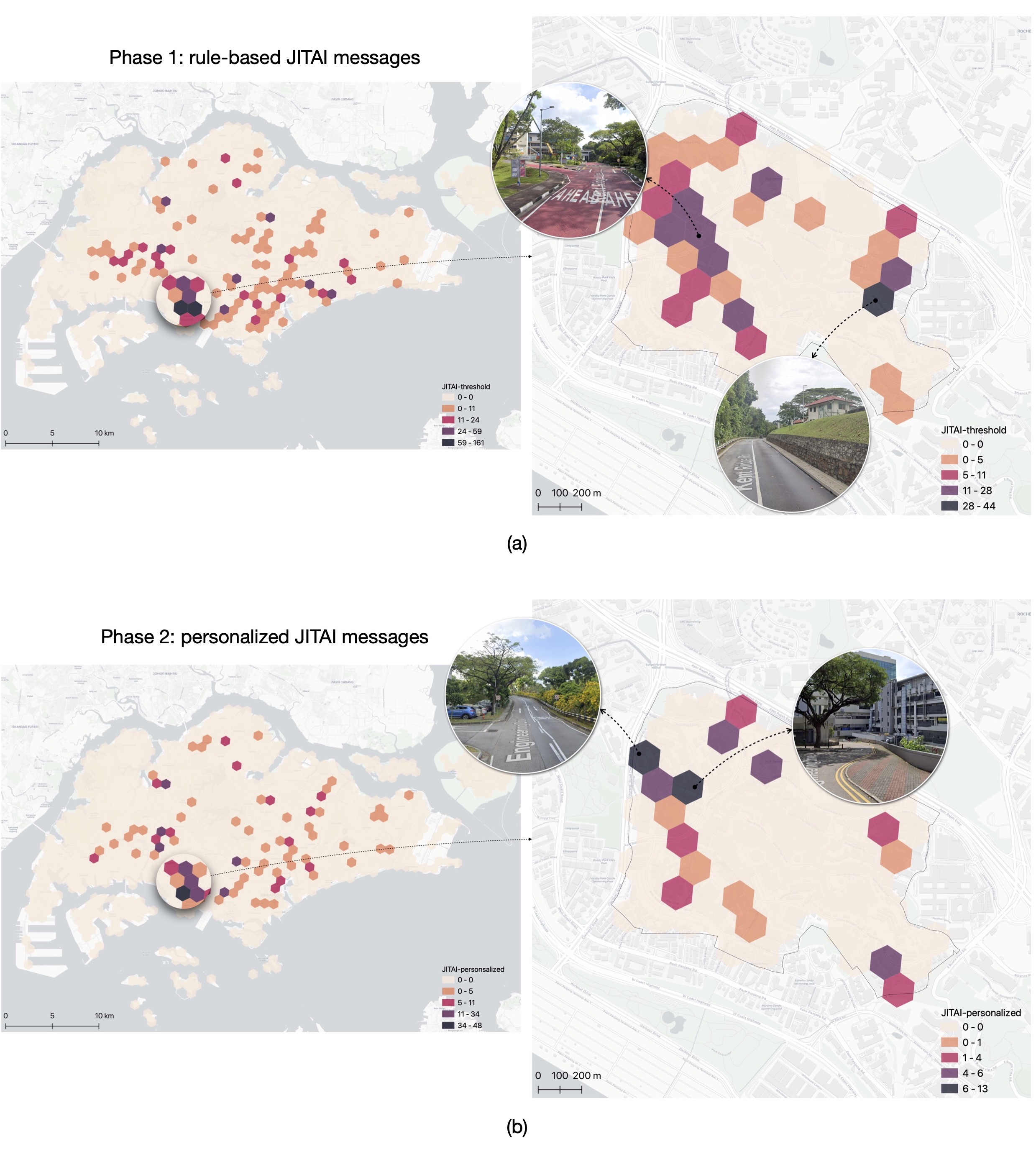}
    \caption{The spatial distribution of JITAI messages nudged in two phases: (a) Phase 1, based on rule-based triggers using outdoor air temperature threshold; and (b) Phase 2, personalized by individual preference. The city-wide distribution demonstrates intervention density across Singapore, while the localized example illustrates detailed spatial differences between the two phases. The color scale represents the number of interventions per spatial unit (hexagon). Basemap and imagery: (c) OpenStreetMap contributors and Google Street View.}
    \label{fig:jitai-spatial}
\end{figure*}

Figure~\ref{fig:jitai-spatial} visualizes the rule-based JITAI messages (Phase 1) and personalized prediction triggers (Phase 2) with respect to the spatial distribution of JITAI notifications. In Phase 1, the rule-based JITAI, determined solely by outdoor temperature thresholds, is mapped. The visualization of Phase 1 illustrates that most notifications are distributed along the southwest part of Singapore, with several dense clusters scattered (e.g., the industrially diverse Jurong area). A detailed examination of a university campus that many participants regularly visited reveals a notably dense distribution of messages, comprising 16.4\% of threshold messages. A spatial distribution of rule-based messages indicates specific areas on campus where participants are most likely to receive interventions, alerting to unsatisfied comfort. Theoretically, outdoor locations near main roads may result in warmer comfort compared to semi-open or sheltered areas in tropical weather.
Phase 2 incorporates individual preferences to generate personalized JITAI prediction messages. The spatial visualization in Singapore is consistent with Phase 1, but shows a more compact distribution in particular areas. Meanwhile, a closer look at campus data suggests that personalized adjustments enable more accurate locations for sending messages, for example, pedestrian paths without shading or with steep slopes. This spatially explicit information implies that participants might experience warmer comfort in these targeted areas.

\section{Discussion}

The deployment of JITAI in the urban context exposed evidence of the potential effectiveness of this technique in the built environment. Despite the potential, there are several limitations to this deployment and opportunities for exploration. This section discusses the JITAI components and design principles in a framework widely accepted in the mobile health community \citep{Nahum-Shani2018-hw}.

\subsection{Individualization and adaptation}

The primary purpose of JITAI is to provide behavior change support at the right time while minimizing interventions that are interruptive or not beneficial. 
The noise and thermal comfort context in indoor and outdoor spaces still has many opportunities to explore when the best time to provide intervention messages, what those messages should contain, and which people would be amenable to those messages.
This study focused on two relatively simple methods for triggering and delivering messages. 
All participants in this study received the same interventions and no different combinations of message content were tested. 
Future research should focus on testing a variety of message types with diverse content to evaluate the effectiveness of different strategies.

\begin{figure*}
    \centering
    \begin{subfigure}{\textwidth}
         \includegraphics[width=\textwidth]{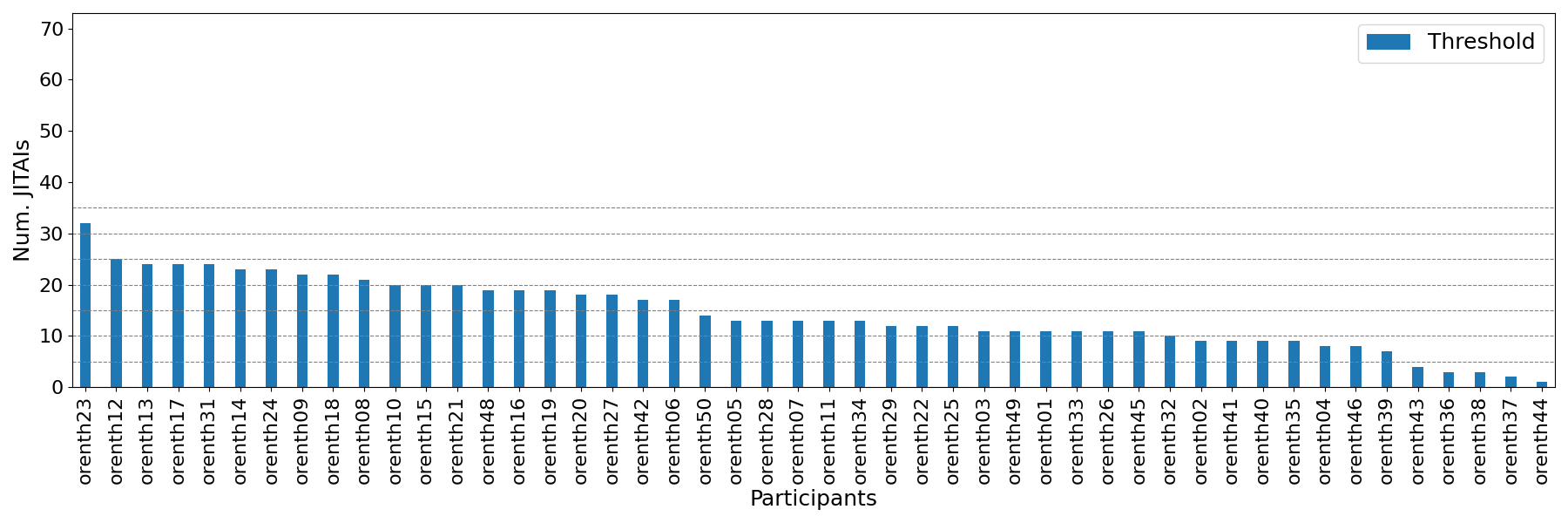}
         \caption{
         Phase 1: Threshold-based JITAIs were only sent after the participant submitted 50 micro-surveys.
         No personalized JITAIs were used in this phase.
         }
         \label{fig:orenth_before_after_jitai}
     \end{subfigure}
     \vfill
     \centering
     \begin{subfigure}{\textwidth}
         \includegraphics[width=\textwidth]{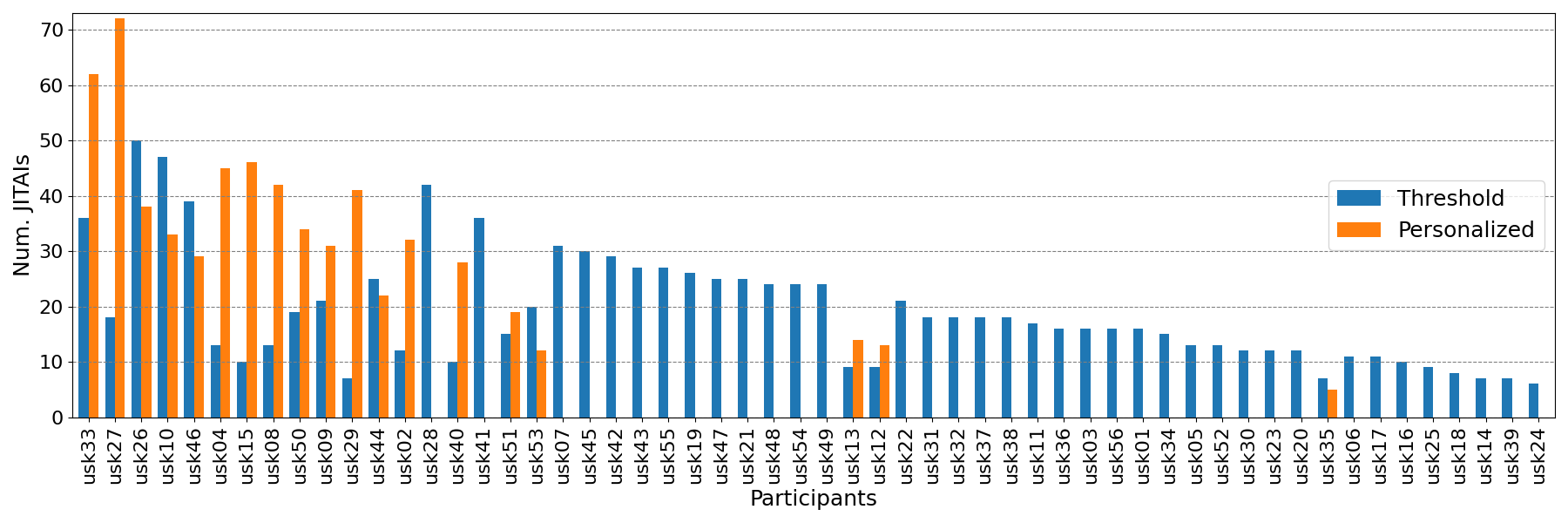}
        \caption{
        Phase 2: Threshold-based JITAIs were only sent until the participant submitted 50 micro-surveys; after that, personalized-based JITAIs were sent until 50 more micro-surveys were submitted.
        }
        \label{fig:usk_before_after_jitai}
     \end{subfigure}
    \caption{
        Total number of JITAIs sent to each participant for all participants in Phase 1 and Phase 2.
    }
    \label{fig:before_after_jitai}
\end{figure*}

Regarding intervention delivery, Figure \ref{fig:before_after_jitai} shows an overview of the number of total JITAI messages sent to participants during the two phases of the experiment.
Figure \ref{fig:orenth_before_after_jitai} shows that when JITAIs are activated by a fixed threshold, all participants receive at least a couple of messages.
The same situation is seen for the other set of participants in Figure \ref{fig:usk_before_after_jitai}.
However, when JITAI messages are triggered by a personalized mechanism, only some participants receive these messages.
Figure \ref{fig:usk_before_after_jitai} shows how only 19 of the 55 participants received these personalized messages (orange bars).
Within this group, only three participants received a total of fewer personalized JITAI messages than threshold-based messages, demonstrating the sensitivity of personalization.
In addition, there are numerous opportunities to improve the use of interventions in combination with building controls \citep{Lorenz2023-ok}, personal comfort models \citep{Arakawa_Martins2022-vh}, and other human-building interaction strategies \citep{Becerik-Gerber2022-am}.

\subsection{State of vulnerability and opportunity}

Within the JITAI community focused on health outcomes, there is a focus on identifying periods of susceptibility to negative or positive effects. These periods are crucial in determining the best context and instance for providing information. In the context of the built environment, these are the instances in which someone would be impacted by their environmental conditions to the point of degradation of their health, productivity, or satisfaction state. There are numerous opportunities to design experiments that can better identify these states for thermal and aural comfort. In this study, the key focus of the intervention design was based on assumptions of when environmental variables would cross various thresholds. Future built environment studies utilizing JITAI should investigate more detailed and sophisticated means of detecting the right context for intervention.

\subsection{Decision points and tailoring variables}

Within an intervention deployment, there are specific triggers that are used to send the message at the right time. These situations are known as decision points, and the data that are used to target the right time for a specific person are called tailoring variables that may require adaptation to local contexts, e.g., outside air temperature thresholds would change based on climate and seasons. In this deployment, sound levels and outdoor air temperature were used as the tailoring variables. The findings offer a nuanced understanding of the factors that influence the perception and efficacy of JITAI messages in an urban context. The data suggest that a multifaceted approach is necessary when designing JITAIs, one that accounts for individual differences in sensitivity to stimuli, life satisfaction, and environmental preferences. Tailoring messages to align with these individual characteristics can improve the acceptance and effectiveness of interventions. Future research should explore the development of customizable JITAIs that can adapt in real-time to users' feedback, maximizing comfort, and minimizing disruption. Practical applications of this research could lead to more intelligent and user-centered environmental control systems in workplaces, ultimately promoting well-being and productivity. The potential of JITAIs to improve indoor environmental quality is significant, but it depends on our ability to integrate individual human factors into their design and implementation.

Another fault of the implemented framework is that the JITAI interventions were formulated at a relatively high level and with an unsophisticated view of noise and thermal comfort. The high-level view is forced in part by the limitations of the standard sensor hardware used in this study. The noise measurement provided by the Apple Watch cannot distinguish different sources of noise. In addition, water and wind can degrade the accuracy of the measurement.\footnote{\url{http://support.apple.com/en-sg/102315}}
There is a lot of opportunity to investigate more deeply the caveats of how to nudge people to change their behavior in these aspects. For example, the concept of noise in the urban context is not just a matter of measuring sound pressure level, as there are more complex frameworks of analysis that can be used for evaluation and behavior change interventions.


\subsection{Role of urban-scale digital twins and spatial information}
Given the momentum of the JITAI framework in providing personalized information in the context of the built environment~\citep{Hardeman2019-gs,reichert2020studying,ligtenberg2022making,tobin2023use}, the analysis can be expanded to an urban scale, integrating with geospatial data. As analyzed in Section~\ref{sec:spatial}, the distribution of JITAI interventions shows a geospatial heterogeneity related to various surroundings, e.g., shedding areas, vegetation coverage, and slope of the road. Such spatial aspects play a role in measuring and representing the urban environment to facilitate subsequent studies in various domains, for example, understanding thermal environments~\citep{yang2021influence,wang20243d}. Exploiting geospatial information, the JITAI framework can be further adapted to enable spatial triggers or geo-fencing, which are discussed as live tracking for real-time activities in multiple disciplines, e.g. disaster management, bike sharing planning, and mobility \citep{szczytowski2014geo,gupta2016child,mangold2022geo}. Considering dynamic and explicit spatial differences, it uncovers the potential to quantify context-specific information and generate geofencing, notifying people of their urban surroundings regarding their locations. For example, analyzing the portion of urban features from geotagged images such as vegetation, sky, and buildings can be useful to assess outdoor comfort~\citep{2023_scs_human_dt,2023_npjus_urbanity}.

Meanwhile, the JITAI framework, as an example of a two-way interaction between humans and the urban environment, innovates a void of a human perspective in urban digital twins and completes the information loop among people, urban systems, and virtual models. An urban digital twin is an advanced approach to dynamically simulating urban scenarios and managing city development by modeling and replicating physical environments and their processes~\citep{Lehtola2022}. However, the current discourse on urban digital twins is primarily driven by data and techniques, yet overlooks social and human aspects when operating and adopting them in practice~\citep{2023_autcon_dt_challenges}. The JITAI framework provides valuable information on the collection of human feedback and the triggering of personalized interventions. Therefore, it suggests future deployment to fulfill the role of humans in urban digital twins and, meanwhile, to spur the notion of dynamic and seamless information exchange between physical and virtual systems. For example, advanced by urban digital twins, the JITAI approach can be adapted to infer how people perceive their neighborhoods, thus improving livability, using an amalgamation of spatial information (e.g., urban elements), demographic characteristics (e.g., age), and individual perceptions. 

\subsection{Proximal outcomes}

The proximal outcomes of an intervention are the immediate behavioral changes made by a person as a result of the messages. In this study, they were related to changing location or using headphones to mitigate environmental issues. The use of headphones may also be interpreted through the lens of acoustic Personal Environmental Control Systems (PECS), which emphasize individual agency in the management of auditory comfort through wearable technologies \citep{Torresin2024PECS}. The results of this study underscore the intricate dynamics of human responses to environmental changes, especially within the framework of Just-In-Time Adaptive Intervention (JITAI) systems. Participants demonstrated a willingness to adapt their behaviors in response to interventions that target discomfort, exhibiting varying degrees of responsiveness throughout the duration of the study. In particular, Phase 2 provided compelling evidence of proactive behavioral adjustments, affirming the efficacy of the intervention messages. These results not only reinforce the importance of JITAI systems in facilitating adaptive human-environment interactions but also suggest avenues for refining these systems to enhance their impact. Future research may benefit from exploring the longitudinal effects of such interventions and their integration into daily life to maintain sustained behavioral change. However, a closer examination of the response patterns reveals that the participants may have found it easier or more habitual to respond to noise-related discomfort than to thermal discomfort. Acoustic interventions showed a consistent increase in behavioral adoption (for example, location change and use of earphones), while thermal intervention responses remained relatively stable after the first week. 

\subsection{Distal outcomes}

Distal outcomes are the long-term objectives of the JITAI interventions. For the built environment, these are related to the improvement in productivity and satisfaction over time due to behavior change. Although this study focused on proximal outcomes (that is, immediate behavioral responses), it did not assess whether these changes led to actual improvements in comfort, well-being, or performance. This was outside the scope and time frame of the study. This whole category of effects could be a target for future studies that incorporate a longer time range, specific subjective or objective measurements of productivity or health, and innovations in the intervention and micro-survey content.

\subsection{Intervention engagement and fatigue}
For many people, there is a fine line between interruptions that help improve their situation and annoyance in being bothered. In this study, there was a weekly question to collect a subjective evaluation of the level of annoyance caused by the intervention messages. The results showed that half or more of the participants found some aspect of this feeling and the intensity increased as the study progressed. Phase 2 deployment had an intervention trigger that was more sensitive for some participants, resulting in increased messages and higher levels of annoyance on average. There is a significant amount of potential work to be done in the area of finding the right balance of intervention timings and targeting to mitigate distraction and fatigue. In addition, there is evidence from this study that JITAI as a method could be inappropriate for some people based on their personal characteristics such as age and personality attributes. Furthermore, one limitation of the current analysis is that we did not control for potential changes in environmental conditions (e.g., noise levels or outdoor temperatures) over weeks. As such, we cannot definitively determine whether the increased behavioral responses over time were driven by exposure to the intervention alone or by changes in ambient discomfort. This limitation reflects the challenges of field-based studies, where environmental variability is an inherent feature of real-world deployment.

\subsection{Evaluation of effectiveness and outcomes}
This study relied on subjective surveys to evaluate the effectiveness and \emph{usefulness} of the JITAI method in influencing behavior. A major fault of the implemented method was the inability of the insights to quantify the actual magnitude of impact on the behavior of the participants. The framework deployed lacked measurable observation that would provide objective evidence of a behavior change. It is suggested that future researchers create this kind of structure in their method and provide a means of controlled comparison in which participants do not receive intervention messages to test the effectiveness of the method.

\section{Conclusion}
This paper outlines one of the first implementations of Just-in-Time Adaptive Interventions (JITAI) as a method in the built environment thermal and aural comfort context. The deployment of a smartwatch-based methodology in 103 participants in indoor and outdoor environments was completed and analyzed. The results shed light on the usefulness of this method in the context of identifying human-centric ways to mitigate the impact of heat and noise. Two different intervention delivery methods were tested in two phases. The results of the weekly survey for the first three weeks showed an overall increase in perceived usefulness ranging from 8-19\%. For noise-related interventions, participants demonstrated an increase in location changes ranging from 4-11\% and a 2-17\% increase in the use of earphones to mitigate noise distractions. For interventions related to thermal comfort, participants showed a 3-13\% increase in adjustments to their location or thermostat to improve comfort. The analysis also shows that personality traits (such as conscientiousness), gender, and environmental preferences are key factors in determining the perceived helpfulness of JITAIs, as well as in influencing behavior change. However, the study also raised many questions about the types of people who would find these interventions useful or not and the long-term distal outcomes of such a method. There is evidence that the tolerance of such a method fades over time, and there is significant work to be done to explore the balance of distraction and usefulness.

\subsection{Reproducibility}
All datasets, scripts, and analysis code are openly available in a GitHub repository\footnote{\url{https://github.com/buds-lab/make-yourself-comfortable-jitai-journal-paper}} to support transparency and reproducibility. Documentation within the repository outlines the analysis workflow, enabling verification and further exploration of findings. Regression and chi-square analyzes are included in the data set for an in-depth examination of the associations of variables.

\section*{CRediT author statement}
\textbf{CM: }
Conceptualization,
Methodology,
Investigation,
Resources,
Writing - Original Draft,
Visualization,
Supervision,
Project administration,
Funding acquisition;
\textbf{YXC: }
Conceptualization,
Methodology,
Software,
Validation,
Formal analysis,
Investigation,
Data Curation,
Writing - Review \& Editing,
Visualization,
Project administration;
\textbf{MQ: }
Conceptualization,
Methodology,
Software,
Validation,
Investigation,
Data Curation,
Writing - Review \& Editing
;
\textbf{BL: }
Software,
Formal analysis,
Writing - Review \& Editing,
Visualization
;
\textbf{FB: }
Formal analysis,
Resources,
Writing - Review \& Editing,
Visualization,
Supervision,
Funding acquisition;
\textbf{MF: }
Conceptualization,
Methodology,
Software,
Validation,
Investigation,
Data Curation,
Writing - Review \& Editing,
Visualization,
Project administration.

\section*{Funding}
This research was supported by the following Singapore Ministry of Education (MOE) Tier 1 Grants: The Internet-of-Buildings (IoB) Platform – Testing of Visual Analytics for AI Technologies towards a Well and Green Built Environment (A-0008305-01-00), Ecological Momentary Assessment (EMA) for Built Environment Research (A-0008301-01-00), and Multiscale Digital Twins for the Urban Environment: From Heartbeats to Cities (A-8000139-01-00). This research was secondarily supported by the National Research Foundation, Prime Minister’s Office, Singapore, under its Campus for Research Excellence and Technological Enterprise (CREATE) programme.

\section*{Acknowledgments}
The authors thank Andre Matthias Mueller from the NUS School of Public Health for discussions in the early phases of protocol development. The authors acknowledge the following student researchers who assisted in data collection and processing, including Charis Boey, Kristi Maisha, Annabel Sim, Aisyah Iskandar, Lum Jun Chao, and S~Aravindkugesh.

 \bibliographystyle{elsarticle-harv} 
 \bibliography{references}

\begin{thebibliography}{57}
\expandafter\ifx\csname natexlab\endcsname\relax\def\natexlab#1{#1}\fi
\providecommand{\url}[1]{\texttt{#1}}
\providecommand{\href}[2]{#2}
\providecommand{\path}[1]{#1}
\providecommand{\DOIprefix}{doi:}
\providecommand{\ArXivprefix}{arXiv:}
\providecommand{\URLprefix}{URL: }
\providecommand{\Pubmedprefix}{pmid:}
\providecommand{\doi}[1]{\href{http://dx.doi.org/#1}{\path{#1}}}
\providecommand{\Pubmed}[1]{\href{pmid:#1}{\path{#1}}}
\providecommand{\bibinfo}[2]{#2}
\ifx\xfnm\relax \def\xfnm[#1]{\unskip,\space#1}\fi
\bibitem[{Abboushi et~al.(2022)Abboushi, Safranek, Rodriguez-Feo~Bermudez,
  Pratoomratana, Chen, Poplawski and Davis}]{Abboushi2022-ot}
\bibinfo{author}{Abboushi, B.}, \bibinfo{author}{Safranek, S.},
  \bibinfo{author}{Rodriguez-Feo~Bermudez, E.}, \bibinfo{author}{Pratoomratana,
  S.}, \bibinfo{author}{Chen, Y.}, \bibinfo{author}{Poplawski, M.},
  \bibinfo{author}{Davis, R.}, \bibinfo{year}{2022}.
\newblock \bibinfo{title}{A review of the use of wearables in indoor
  environmental quality studies and an evaluation of data accessibility from a
  wearable device}.
\newblock \bibinfo{journal}{Frontiers in Built Environment}
  \bibinfo{volume}{8}.
\bibitem[{Al{\'\i}as and Alsina-Pag{\`e}s(2019)}]{Alias2019-le}
\bibinfo{author}{Al{\'\i}as, F.}, \bibinfo{author}{Alsina-Pag{\`e}s, R.M.},
  \bibinfo{year}{2019}.
\newblock \bibinfo{title}{Review of wireless acoustic sensor networks for
  environmental noise monitoring in smart cities}.
\newblock \bibinfo{journal}{Journal of Sensors} \bibinfo{volume}{2019}.
\bibitem[{Angelici and Profeta(2023)}]{Angelici2023-rw}
\bibinfo{author}{Angelici, M.}, \bibinfo{author}{Profeta, P.},
  \bibinfo{year}{2023}.
\newblock \bibinfo{title}{Smart working: Work flexibility without constraints}.
\newblock \bibinfo{journal}{Manage. Sci.} .
\bibitem[{Appel-Meulenbroek et~al.(2020)Appel-Meulenbroek, Steps, Wenmaekers
  and Arentze}]{Appel-Meulenbroek2020-bp}
\bibinfo{author}{Appel-Meulenbroek, R.}, \bibinfo{author}{Steps, S.},
  \bibinfo{author}{Wenmaekers, R.}, \bibinfo{author}{Arentze, T.},
  \bibinfo{year}{2020}.
\newblock \bibinfo{title}{Coping strategies and perceived productivity in
  open-plan offices with noise problems}.
\newblock \bibinfo{journal}{Journal of Managerial Psychology}
  \bibinfo{volume}{36}, \bibinfo{pages}{400--414}.
\bibitem[{Arakawa~Martins et~al.(2022)Arakawa~Martins, Soebarto and
  Williamson}]{Arakawa_Martins2022-vh}
\bibinfo{author}{Arakawa~Martins, L.}, \bibinfo{author}{Soebarto, V.},
  \bibinfo{author}{Williamson, T.}, \bibinfo{year}{2022}.
\newblock \bibinfo{title}{A systematic review of personal thermal comfort
  models}.
\newblock \bibinfo{journal}{Build. Environ.} \bibinfo{volume}{207},
  \bibinfo{pages}{108502}.
\bibitem[{Becerik-Gerber et~al.(2022)Becerik-Gerber, Lucas, Aryal, Awada,
  Berg{\'e}s, Billington, Boric-Lubecke, Ghahramani, Heydarian, H{\"o}elscher,
  Jazizadeh, Khan, Langevin, Liu, Marks, Mauriello, Murnane, Noh, Pritoni,
  Roll, Schaumann, Seyedrezaei, Taylor, Zhao and Zhu}]{Becerik-Gerber2022-am}
\bibinfo{author}{Becerik-Gerber, B.}, \bibinfo{author}{Lucas, G.},
  \bibinfo{author}{Aryal, A.}, \bibinfo{author}{Awada, M.},
  \bibinfo{author}{Berg{\'e}s, M.}, \bibinfo{author}{Billington, S.},
  \bibinfo{author}{Boric-Lubecke, O.}, \bibinfo{author}{Ghahramani, A.},
  \bibinfo{author}{Heydarian, A.}, \bibinfo{author}{H{\"o}elscher, C.},
  \bibinfo{author}{Jazizadeh, F.}, \bibinfo{author}{Khan, A.},
  \bibinfo{author}{Langevin, J.}, \bibinfo{author}{Liu, R.},
  \bibinfo{author}{Marks, F.}, \bibinfo{author}{Mauriello, M.L.},
  \bibinfo{author}{Murnane, E.}, \bibinfo{author}{Noh, H.},
  \bibinfo{author}{Pritoni, M.}, \bibinfo{author}{Roll, S.},
  \bibinfo{author}{Schaumann, D.}, \bibinfo{author}{Seyedrezaei, M.},
  \bibinfo{author}{Taylor, J.E.}, \bibinfo{author}{Zhao, J.},
  \bibinfo{author}{Zhu, R.}, \bibinfo{year}{2022}.
\newblock \bibinfo{title}{The field of human building interaction for
  convergent research and innovation for intelligent built environments}.
\newblock \bibinfo{journal}{Sci. Rep.} \bibinfo{volume}{12},
  \bibinfo{pages}{22092}.
\bibitem[{Bergefurt et~al.(2023)Bergefurt, Appel-Meulenbroek, Maris, Arentze,
  Weijs-Perr{\'e}e and de~Kort}]{Bergefurt2023-fm}
\bibinfo{author}{Bergefurt, L.}, \bibinfo{author}{Appel-Meulenbroek, R.},
  \bibinfo{author}{Maris, C.}, \bibinfo{author}{Arentze, T.},
  \bibinfo{author}{Weijs-Perr{\'e}e, M.}, \bibinfo{author}{de~Kort, Y.},
  \bibinfo{year}{2023}.
\newblock \bibinfo{title}{The influence of distractions of the home-work
  environment on mental health during the {COVID-19} pandemic}.
\newblock \bibinfo{journal}{Ergonomics} \bibinfo{volume}{66},
  \bibinfo{pages}{16--33}.
\bibitem[{Castaldo et~al.(2018)Castaldo, Pigliautile, Rosso, Cotana, De~Giorgio
  and Pisello}]{Castaldo2018-ih}
\bibinfo{author}{Castaldo, V.L.}, \bibinfo{author}{Pigliautile, I.},
  \bibinfo{author}{Rosso, F.}, \bibinfo{author}{Cotana, F.},
  \bibinfo{author}{De~Giorgio, F.}, \bibinfo{author}{Pisello, A.L.},
  \bibinfo{year}{2018}.
\newblock \bibinfo{title}{How subjective and non-physical parameters affect
  occupants' environmental comfort perception}.
\newblock \bibinfo{journal}{Energy Build.} \bibinfo{volume}{178},
  \bibinfo{pages}{107--129}.
\bibitem[{Cheung et~al.(2019)Cheung, Schiavon, Parkinson, Li and
  Brager}]{Cheung2019-fs}
\bibinfo{author}{Cheung, T.}, \bibinfo{author}{Schiavon, S.},
  \bibinfo{author}{Parkinson, T.}, \bibinfo{author}{Li, P.},
  \bibinfo{author}{Brager, G.}, \bibinfo{year}{2019}.
\newblock \bibinfo{title}{Analysis of the accuracy on {PMV} -- {PPD} model
  using the {ASHRAE} global thermal comfort database {II}}.
\newblock \bibinfo{journal}{Build. Environ.} \bibinfo{volume}{153},
  \bibinfo{pages}{205--217}.
\bibitem[{Day et~al.(2020)Day, McIlvennie, Brackley, Tarantini, Piselli, Hahn,
  O'Brien, Rajus, De~Simone, Kj{\ae}rgaard, Pritoni, Schl{\"u}ter, Peng,
  Schweiker, Fajilla, Becchio, Fabi, Spigliantini, Derbas and
  Pisello}]{Day2020-tq}
\bibinfo{author}{Day, J.K.}, \bibinfo{author}{McIlvennie, C.},
  \bibinfo{author}{Brackley, C.}, \bibinfo{author}{Tarantini, M.},
  \bibinfo{author}{Piselli, C.}, \bibinfo{author}{Hahn, J.},
  \bibinfo{author}{O'Brien, W.}, \bibinfo{author}{Rajus, V.S.},
  \bibinfo{author}{De~Simone, M.}, \bibinfo{author}{Kj{\ae}rgaard, M.B.},
  \bibinfo{author}{Pritoni, M.}, \bibinfo{author}{Schl{\"u}ter, A.},
  \bibinfo{author}{Peng, Y.}, \bibinfo{author}{Schweiker, M.},
  \bibinfo{author}{Fajilla, G.}, \bibinfo{author}{Becchio, C.},
  \bibinfo{author}{Fabi, V.}, \bibinfo{author}{Spigliantini, G.},
  \bibinfo{author}{Derbas, G.}, \bibinfo{author}{Pisello, A.L.},
  \bibinfo{year}{2020}.
\newblock \bibinfo{title}{A review of select human-building interfaces and
  their relationship to human behavior, energy use and occupant comfort}.
\newblock \bibinfo{journal}{Build. Environ.} \bibinfo{volume}{178},
  \bibinfo{pages}{106920}.
\bibitem[{Fischer et~al.(2022)Fischer, Schraivogel, Caversaccio and
  Wimmer}]{Fischer2022-lv}
\bibinfo{author}{Fischer, T.}, \bibinfo{author}{Schraivogel, S.},
  \bibinfo{author}{Caversaccio, M.}, \bibinfo{author}{Wimmer, W.},
  \bibinfo{year}{2022}.
\newblock \bibinfo{title}{Are smartwatches a suitable tool to monitor noise
  exposure for public health awareness and otoprotection?}
\newblock \bibinfo{journal}{Front. Neurol.} \bibinfo{volume}{13},
  \bibinfo{pages}{856219}.
\bibitem[{Graham et~al.(2021)Graham, Parkinson and Schiavon}]{Graham2021-en}
\bibinfo{author}{Graham, L.t.}, \bibinfo{author}{Parkinson, T.},
  \bibinfo{author}{Schiavon, S.}, \bibinfo{year}{2021}.
\newblock \bibinfo{title}{Lessons learned from 20 years of {CBE's} occupant
  surveys}.
\newblock \bibinfo{journal}{Buildings and Cities} \bibinfo{volume}{2},
  \bibinfo{pages}{166--184}.
\bibitem[{Gupta and Harit(2016)}]{gupta2016child}
\bibinfo{author}{Gupta, A.}, \bibinfo{author}{Harit, V.}, \bibinfo{year}{2016}.
\newblock \bibinfo{title}{Child safety \& tracking management system by using
  gps, geo-fencing \& android application: An analysis}, in:
  \bibinfo{booktitle}{2016 Second International Conference on Computational
  Intelligence \& Communication Technology (CICT)},
  \bibinfo{organization}{IEEE}. pp. \bibinfo{pages}{683--686}.
\bibitem[{Hardeman et~al.(2019)Hardeman, Houghton, Lane, Jones and
  Naughton}]{Hardeman2019-gs}
\bibinfo{author}{Hardeman, W.}, \bibinfo{author}{Houghton, J.},
  \bibinfo{author}{Lane, K.}, \bibinfo{author}{Jones, A.},
  \bibinfo{author}{Naughton, F.}, \bibinfo{year}{2019}.
\newblock \bibinfo{title}{A systematic review of just-in-time adaptive
  interventions ({JITAIs}) to promote physical activity}.
\newblock \bibinfo{journal}{Int. J. Behav. Nutr. Phys. Act.}
  \bibinfo{volume}{16}, \bibinfo{pages}{31}.
\bibitem[{Heydarian et~al.(2020)Heydarian, McIlvennie, Arpan, Yousefi,
  Syndicus, Schweiker, Jazizadeh, Rissetto, Pisello, Piselli, Berger, Yan and
  Mahdavi}]{Heydarian2020-is}
\bibinfo{author}{Heydarian, A.}, \bibinfo{author}{McIlvennie, C.},
  \bibinfo{author}{Arpan, L.}, \bibinfo{author}{Yousefi, S.},
  \bibinfo{author}{Syndicus, M.}, \bibinfo{author}{Schweiker, M.},
  \bibinfo{author}{Jazizadeh, F.}, \bibinfo{author}{Rissetto, R.},
  \bibinfo{author}{Pisello, A.L.}, \bibinfo{author}{Piselli, C.},
  \bibinfo{author}{Berger, C.}, \bibinfo{author}{Yan, Z.},
  \bibinfo{author}{Mahdavi, A.}, \bibinfo{year}{2020}.
\newblock \bibinfo{title}{What drives our behaviors in buildings? a review on
  occupant interactions with building systems from the lens of behavioral
  theories}.
\newblock \bibinfo{journal}{Build. Environ.} \bibinfo{volume}{179},
  \bibinfo{pages}{106928}.
\bibitem[{Jayathissa et~al.(2019)Jayathissa, Quintana, Sood, Nazarian and
  Miller}]{Jayathissa2019-kg}
\bibinfo{author}{Jayathissa, P.}, \bibinfo{author}{Quintana, M.},
  \bibinfo{author}{Sood, T.}, \bibinfo{author}{Nazarian, N.},
  \bibinfo{author}{Miller, C.}, \bibinfo{year}{2019}.
\newblock \bibinfo{title}{Is your clock-face cozie? a smartwatch methodology
  for the in-situ collection of occupant comfort data}.
\newblock \bibinfo{journal}{J. Phys. Conf. Ser.} \bibinfo{volume}{1343},
  \bibinfo{pages}{012145}.
\bibitem[{Keller et~al.(2022)Keller, Hartmann, Teepe, Lohse, Alattas,
  Tudor~Car, M{\"u}ller-Riemenschneider, von Wangenheim, Mair and
  Kowatsch}]{Keller2022-uo}
\bibinfo{author}{Keller, R.}, \bibinfo{author}{Hartmann, S.},
  \bibinfo{author}{Teepe, G.W.}, \bibinfo{author}{Lohse, K.M.},
  \bibinfo{author}{Alattas, A.}, \bibinfo{author}{Tudor~Car, L.},
  \bibinfo{author}{M{\"u}ller-Riemenschneider, F.}, \bibinfo{author}{von
  Wangenheim, F.}, \bibinfo{author}{Mair, J.L.}, \bibinfo{author}{Kowatsch,
  T.}, \bibinfo{year}{2022}.
\newblock \bibinfo{title}{Digital behavior change interventions for the
  prevention and management of type 2 diabetes: Systematic market analysis}.
\newblock \bibinfo{journal}{J. Med. Internet Res.} \bibinfo{volume}{24},
  \bibinfo{pages}{e33348}.
\bibitem[{Kim and de~Dear(2013)}]{Kim2013-gs}
\bibinfo{author}{Kim, J.}, \bibinfo{author}{de~Dear, R.}, \bibinfo{year}{2013}.
\newblock \bibinfo{title}{Workspace satisfaction: The privacy-communication
  trade-off in open-plan offices}.
\newblock \bibinfo{journal}{J. Environ. Psychol.} \bibinfo{volume}{36},
  \bibinfo{pages}{18--26}.
\bibitem[{Kumar et~al.(2021)Kumar, Sarkar and Chahar}]{Kumar2021-zm}
\bibinfo{author}{Kumar, S.}, \bibinfo{author}{Sarkar, S.},
  \bibinfo{author}{Chahar, B.}, \bibinfo{year}{2021}.
\newblock \bibinfo{title}{A systematic review of work-life integration and role
  of flexible work arrangements}.
\newblock \bibinfo{journal}{Int. J. Organ. Anal.}
  \bibinfo{volume}{ahead-of-print}.
\bibitem[{Lehtola et~al.(2022)Lehtola, Koeva, Elberink, Raposo, Virtanen,
  Vahdatikhaki and Borsci}]{Lehtola2022}
\bibinfo{author}{Lehtola, V.V.}, \bibinfo{author}{Koeva, M.},
  \bibinfo{author}{Elberink, S.O.}, \bibinfo{author}{Raposo, P.},
  \bibinfo{author}{Virtanen, J.P.}, \bibinfo{author}{Vahdatikhaki, F.},
  \bibinfo{author}{Borsci, S.}, \bibinfo{year}{2022}.
\newblock \bibinfo{title}{Digital twin of a city: Review of technology serving
  city needs}.
\newblock \bibinfo{journal}{International Journal of Applied Earth Observation
  and Geoinformation} \bibinfo{volume}{114}, \bibinfo{pages}{102915}.
\newblock \DOIprefix\doi{10.1016/j.jag.2022.102915}.
\bibitem[{Lei et~al.(2023)Lei, Janssen, Stoter and
  Biljecki}]{2023_autcon_dt_challenges}
\bibinfo{author}{Lei, B.}, \bibinfo{author}{Janssen, P.},
  \bibinfo{author}{Stoter, J.}, \bibinfo{author}{Biljecki, F.},
  \bibinfo{year}{2023}.
\newblock \bibinfo{title}{Challenges of urban digital twins: A systematic
  review and a delphi expert survey}.
\newblock \bibinfo{journal}{Automation in Construction} \bibinfo{volume}{147},
  \bibinfo{pages}{104716}.
\newblock \DOIprefix\doi{10.1016/j.autcon.2022.104716}.
\bibitem[{Li et~al.(2021)Li, Loveday and Demian}]{Li2021-ej}
\bibinfo{author}{Li, Z.}, \bibinfo{author}{Loveday, D.},
  \bibinfo{author}{Demian, P.}, \bibinfo{year}{2021}.
\newblock \bibinfo{title}{Nudging and usage of thermal comfort-related
  systems}.
\newblock \bibinfo{journal}{Energy Build.} \bibinfo{volume}{252},
  \bibinfo{pages}{111480}.
\bibitem[{Ligtenberg et~al.(2022)Ligtenberg, Simons, Barhorst and
  Winkens}]{ligtenberg2022making}
\bibinfo{author}{Ligtenberg, A.}, \bibinfo{author}{Simons, M.},
  \bibinfo{author}{Barhorst, M.}, \bibinfo{author}{Winkens, L.},
  \bibinfo{year}{2022}.
\newblock \bibinfo{title}{Making space a better place: Just in time adaptive
  interventions for healthy lifestyles}.
\newblock \bibinfo{journal}{AGILE: GIScience Series} \bibinfo{volume}{3},
  \bibinfo{pages}{45}.
\bibitem[{Liu et~al.(2023)Liu, Zhao, Luo, Lei, Frei, Miller and
  Biljecki}]{2023_scs_human_dt}
\bibinfo{author}{Liu, P.}, \bibinfo{author}{Zhao, T.}, \bibinfo{author}{Luo,
  J.}, \bibinfo{author}{Lei, B.}, \bibinfo{author}{Frei, M.},
  \bibinfo{author}{Miller, C.}, \bibinfo{author}{Biljecki, F.},
  \bibinfo{year}{2023}.
\newblock \bibinfo{title}{Towards human-centric digital twins: Leveraging
  computer vision and graph models to predict outdoor comfort}.
\newblock \bibinfo{journal}{Sustainable Cities and Society}
  \bibinfo{volume}{93}, \bibinfo{pages}{104480}.
\newblock \DOIprefix\doi{10.1016/j.scs.2023.104480}.
\bibitem[{Lorenz et~al.(2023)Lorenz, Andr{\'e}, Abele, Gunay, Hahn, Hensen,
  Nagy, Ouf, Park, Yaduvanshi and Miller}]{Lorenz2023-ok}
\bibinfo{author}{Lorenz, C.L.}, \bibinfo{author}{Andr{\'e}, M.},
  \bibinfo{author}{Abele, O.}, \bibinfo{author}{Gunay, B.},
  \bibinfo{author}{Hahn, J.}, \bibinfo{author}{Hensen, P.},
  \bibinfo{author}{Nagy, Z.}, \bibinfo{author}{Ouf, M.M.},
  \bibinfo{author}{Park, J.Y.}, \bibinfo{author}{Yaduvanshi, N.S.},
  \bibinfo{author}{Miller, C.}, \bibinfo{year}{2023}.
\newblock \bibinfo{title}{A repository of occupant-centric control case
  studies: Survey development and database overview}.
\newblock \bibinfo{journal}{Energy Build.} \bibinfo{volume}{300},
  \bibinfo{pages}{113649}.
\bibitem[{Maisha et~al.(2023)Maisha, Frei, Quintana, Chua, Jain and
  Miller}]{Maisha2023-dt}
\bibinfo{author}{Maisha, K.}, \bibinfo{author}{Frei, M.},
  \bibinfo{author}{Quintana, M.}, \bibinfo{author}{Chua, Y.X.},
  \bibinfo{author}{Jain, R.}, \bibinfo{author}{Miller, C.},
  \bibinfo{year}{2023}.
\newblock \bibinfo{title}{Utilizing wearable technology to characterize and
  facilitate occupant collaborations in flexible workspaces}.
\newblock \bibinfo{journal}{J. Phys. Conf. Ser.} \bibinfo{volume}{2600},
  \bibinfo{pages}{142009}.
\bibitem[{Mangold et~al.(2022)Mangold, Zhao, Haitao and
  Mansourian}]{mangold2022geo}
\bibinfo{author}{Mangold, M.}, \bibinfo{author}{Zhao, P.},
  \bibinfo{author}{Haitao, H.}, \bibinfo{author}{Mansourian, A.},
  \bibinfo{year}{2022}.
\newblock \bibinfo{title}{Geo-fence planning for dockless bike-sharing systems:
  A gis-based multi-criteria decision analysis framework}.
\newblock \bibinfo{journal}{Urban Informatics} \bibinfo{volume}{1},
  \bibinfo{pages}{17}.
\bibitem[{Miller et~al.(2021)Miller, Abdelrahman, Chong, Biljecki, Quintana,
  Frei, Chew and Wong}]{Miller2021-bm}
\bibinfo{author}{Miller, C.}, \bibinfo{author}{Abdelrahman, M.},
  \bibinfo{author}{Chong, A.}, \bibinfo{author}{Biljecki, F.},
  \bibinfo{author}{Quintana, M.}, \bibinfo{author}{Frei, M.},
  \bibinfo{author}{Chew, M.}, \bibinfo{author}{Wong, D.}, \bibinfo{year}{2021}.
\newblock \bibinfo{title}{The {Internet-of-Buildings} ({IoB}) --- digital twin
  convergence of wearable and {IoT} data with {GIS/BIM}}.
\newblock \bibinfo{journal}{J. Phys. Conf. Ser.} \bibinfo{volume}{2042},
  \bibinfo{pages}{012041}.
\bibitem[{Miller et~al.(2022)Miller, Chua, Frei and Quintana}]{Miller2022-dy}
\bibinfo{author}{Miller, C.}, \bibinfo{author}{Chua, Y.X.},
  \bibinfo{author}{Frei, M.}, \bibinfo{author}{Quintana, M.},
  \bibinfo{year}{2022}.
\newblock \bibinfo{title}{Towards smartwatch-driven just-in-time adaptive
  interventions ({JITAI}) for building occupants}, in: \bibinfo{booktitle}{The
  9th {ACM} International Conference on Systems for {Energy-Efficient}
  Buildings, Cities, and Transportation}.
\bibitem[{Miller et~al.(2023)Miller, Quintana, Frei, Chua, Fu, Picchetti, Yap,
  Chong and Biljecki}]{Miller2023-wi}
\bibinfo{author}{Miller, C.}, \bibinfo{author}{Quintana, M.},
  \bibinfo{author}{Frei, M.}, \bibinfo{author}{Chua, Y.X.},
  \bibinfo{author}{Fu, C.}, \bibinfo{author}{Picchetti, B.},
  \bibinfo{author}{Yap, W.}, \bibinfo{author}{Chong, A.},
  \bibinfo{author}{Biljecki, F.}, \bibinfo{year}{2023}.
\newblock \bibinfo{title}{Introducing the cool, quiet city competition:
  Predicting smartwatch-reported heat and noise with digital twin metrics}, in:
  \bibinfo{booktitle}{Proceedings of the 10th ACM International Conference on
  Systems for Energy-Efficient Buildings, Cities, and Transportation},
  \bibinfo{publisher}{Association for Computing Machinery},
  \bibinfo{address}{New York, NY, USA}. pp. \bibinfo{pages}{298--299}.
\bibitem[{Mosteiro-Romero et~al.(2024)Mosteiro-Romero, Park and
  Miller}]{Mosteiro-Romero2024-eb}
\bibinfo{author}{Mosteiro-Romero, M.}, \bibinfo{author}{Park, Y.},
  \bibinfo{author}{Miller, C.}, \bibinfo{year}{2024}.
\newblock \bibinfo{title}{Converging smartwatch and urban datasets for
  sustainable city planning: A case study in seoul, south korea}.
\newblock \bibinfo{journal}{E3S Web Conf.} .
\bibitem[{Mosteiro-Romero et~al.(2023)Mosteiro-Romero, Quintana, Miller and
  Stouffs}]{Mosteiro-Romero2023-us}
\bibinfo{author}{Mosteiro-Romero, M.}, \bibinfo{author}{Quintana, M.},
  \bibinfo{author}{Miller, C.}, \bibinfo{author}{Stouffs, R.},
  \bibinfo{year}{2023}.
\newblock \bibinfo{title}{From personal comfort to district performance: Using
  smartwatch and {WiFi} data for occupant-driven operation}, in:
  \bibinfo{booktitle}{Proceedings of the 10th ACM International Conference on
  Systems for Energy-Efficient Buildings, Cities, and Transportation},
  \bibinfo{publisher}{Association for Computing Machinery},
  \bibinfo{address}{New York, NY, USA}. pp. \bibinfo{pages}{278--279}.
\bibitem[{M{\"u}ller et~al.(2017)M{\"u}ller, Blandford and
  Yardley}]{Muller2017-ao}
\bibinfo{author}{M{\"u}ller, A.M.}, \bibinfo{author}{Blandford, A.},
  \bibinfo{author}{Yardley, L.}, \bibinfo{year}{2017}.
\newblock \bibinfo{title}{The conceptualization of a {Just-In-Time} adaptive
  intervention ({JITAI}) for the reduction of sedentary behavior in older
  adults}.
\newblock \bibinfo{journal}{Mhealth} \bibinfo{volume}{3}, \bibinfo{pages}{37}.
\bibitem[{Nahum-Shani et~al.(2018)Nahum-Shani, Smith, Spring, Collins,
  Witkiewitz, Tewari and Murphy}]{Nahum-Shani2018-hw}
\bibinfo{author}{Nahum-Shani, I.}, \bibinfo{author}{Smith, S.N.},
  \bibinfo{author}{Spring, B.J.}, \bibinfo{author}{Collins, L.M.},
  \bibinfo{author}{Witkiewitz, K.}, \bibinfo{author}{Tewari, A.},
  \bibinfo{author}{Murphy, S.A.}, \bibinfo{year}{2018}.
\newblock \bibinfo{title}{{Just-in-Time} adaptive interventions ({JITAIs}) in
  mobile health: Key components and design principles for ongoing health
  behavior support}.
\newblock \bibinfo{journal}{Ann. Behav. Med.} \bibinfo{volume}{52},
  \bibinfo{pages}{446--462}.
\bibitem[{Parkinson et~al.(2023)Parkinson, Schiavon, Kim and
  Betti}]{Parkinson2023-yk}
\bibinfo{author}{Parkinson, T.}, \bibinfo{author}{Schiavon, S.},
  \bibinfo{author}{Kim, J.}, \bibinfo{author}{Betti, G.}, \bibinfo{year}{2023}.
\newblock \bibinfo{title}{Common sources of occupant dissatisfaction with
  workspace environments in 600 office buildings}.
\newblock \bibinfo{journal}{Build. Cities} \bibinfo{volume}{4},
  \bibinfo{pages}{17--35}.
\bibitem[{Pollard et~al.(2022)Pollard, Engelen, Held, Van~Buskirk, Spinney and
  de~Dear}]{Pollard2022-gl}
\bibinfo{author}{Pollard, B.}, \bibinfo{author}{Engelen, L.},
  \bibinfo{author}{Held, F.}, \bibinfo{author}{Van~Buskirk, J.},
  \bibinfo{author}{Spinney, R.}, \bibinfo{author}{de~Dear, R.},
  \bibinfo{year}{2022}.
\newblock \bibinfo{title}{Associations between spatial attributes, {IEQ}
  exposures and occupant movement behaviour in an open-plan office}.
\newblock \bibinfo{journal}{Build. Environ.} \bibinfo{volume}{212},
  \bibinfo{pages}{108812}.
\bibitem[{Quintana et~al.(2021)Quintana, Abdelrahman, Frei, Tartarini and
  Miller}]{Quintana2021-ka}
\bibinfo{author}{Quintana, M.}, \bibinfo{author}{Abdelrahman, M.},
  \bibinfo{author}{Frei, M.}, \bibinfo{author}{Tartarini, F.},
  \bibinfo{author}{Miller, C.}, \bibinfo{year}{2021}.
\newblock \bibinfo{title}{Longitudinal personal thermal comfort preference data
  in the wild}, in: \bibinfo{booktitle}{Proceedings of the 19th {ACM}
  Conference on Embedded Networked Sensor Systems},
  \bibinfo{publisher}{Association for Computing Machinery},
  \bibinfo{address}{New York, NY, USA}. pp. \bibinfo{pages}{556--559}.
\bibitem[{Quintana et~al.(2022)Quintana, Schiavon, Tartarini, Kim and
  Miller}]{Quintana.2023}
\bibinfo{author}{Quintana, M.}, \bibinfo{author}{Schiavon, S.},
  \bibinfo{author}{Tartarini, F.}, \bibinfo{author}{Kim, J.},
  \bibinfo{author}{Miller, C.}, \bibinfo{year}{2022}.
\newblock \bibinfo{title}{{Cohort comfort models — Using occupant’s
  similarity to predict personal thermal preference with less data}}.
\newblock \bibinfo{journal}{Building and Environment} \bibinfo{volume}{227},
  \bibinfo{pages}{109685}.
\newblock \DOIprefix\doi{10.1016/j.buildenv.2022.109685},
  \href{http://arxiv.org/abs/2208.03078}{{\tt arXiv:2208.03078}}.
\bibitem[{Reichert et~al.(2020)Reichert, Braun, Lautenbach, Zipf,
  Ebner-Priemer, Tost and Meyer-Lindenberg}]{reichert2020studying}
\bibinfo{author}{Reichert, M.}, \bibinfo{author}{Braun, U.},
  \bibinfo{author}{Lautenbach, S.}, \bibinfo{author}{Zipf, A.},
  \bibinfo{author}{Ebner-Priemer, U.}, \bibinfo{author}{Tost, H.},
  \bibinfo{author}{Meyer-Lindenberg, A.}, \bibinfo{year}{2020}.
\newblock \bibinfo{title}{Studying the impact of built environments on human
  mental health in everyday life: methodological developments, state-of-the-art
  and technological frontiers}.
\newblock \bibinfo{journal}{Current opinion in psychology}
  \bibinfo{volume}{32}, \bibinfo{pages}{158--164}.
\bibitem[{Rupp et~al.(2022)Rupp, Parkinson, Kim, Toftum and
  de~Dear}]{Rupp2022-lq}
\bibinfo{author}{Rupp, R.F.}, \bibinfo{author}{Parkinson, T.},
  \bibinfo{author}{Kim, J.}, \bibinfo{author}{Toftum, J.},
  \bibinfo{author}{de~Dear, R.}, \bibinfo{year}{2022}.
\newblock \bibinfo{title}{The impact of occupant's thermal sensitivity on
  adaptive thermal comfort model}.
\newblock \bibinfo{journal}{Build. Environ.} \bibinfo{volume}{207},
  \bibinfo{pages}{108517}.
\bibitem[{Salim et~al.(2020)Salim, Dong, Ouf, Wang, Pigliautile, Kang, Hong,
  Wu, Liu, Rumi, Rahaman, An, Deng, Shao, Dziedzic, Sangogboye, Kj{\ae}rgaard,
  Kong, Fabiani, Pisello and Yan}]{Salim2020-ec}
\bibinfo{author}{Salim, F.D.}, \bibinfo{author}{Dong, B.},
  \bibinfo{author}{Ouf, M.}, \bibinfo{author}{Wang, Q.},
  \bibinfo{author}{Pigliautile, I.}, \bibinfo{author}{Kang, X.},
  \bibinfo{author}{Hong, T.}, \bibinfo{author}{Wu, W.}, \bibinfo{author}{Liu,
  Y.}, \bibinfo{author}{Rumi, S.K.}, \bibinfo{author}{Rahaman, M.S.},
  \bibinfo{author}{An, J.}, \bibinfo{author}{Deng, H.}, \bibinfo{author}{Shao,
  W.}, \bibinfo{author}{Dziedzic, J.}, \bibinfo{author}{Sangogboye, F.C.},
  \bibinfo{author}{Kj{\ae}rgaard, M.B.}, \bibinfo{author}{Kong, M.},
  \bibinfo{author}{Fabiani, C.}, \bibinfo{author}{Pisello, A.L.},
  \bibinfo{author}{Yan, D.}, \bibinfo{year}{2020}.
\newblock \bibinfo{title}{Modelling urban-scale occupant behaviour, mobility,
  and energy in buildings: A survey}.
\newblock \bibinfo{journal}{Build. Environ.} \bibinfo{volume}{183},
  \bibinfo{pages}{106964}.
\bibitem[{Saponaro et~al.(2021)Saponaro, Vemuri, Dominick and
  Decker}]{Saponaro2021-rx}
\bibinfo{author}{Saponaro, M.}, \bibinfo{author}{Vemuri, A.},
  \bibinfo{author}{Dominick, G.}, \bibinfo{author}{Decker, K.},
  \bibinfo{year}{2021}.
\newblock \bibinfo{title}{Contextualization and individualization for
  just-in-time adaptive interventions to reduce sedentary behavior}, in:
  \bibinfo{booktitle}{Proceedings of the Conference on Health, Inference, and
  Learning}, \bibinfo{publisher}{Association for Computing Machinery},
  \bibinfo{address}{New York, NY, USA}. pp. \bibinfo{pages}{246--256}.
\bibitem[{Schweiker et~al.(2020)Schweiker, Rissetto and
  Wagner}]{Schweiker2020-pw}
\bibinfo{author}{Schweiker, M.}, \bibinfo{author}{Rissetto, R.},
  \bibinfo{author}{Wagner, A.}, \bibinfo{year}{2020}.
\newblock \bibinfo{title}{Thermal expectation: Influencing factors and its
  effect on thermal perception}.
\newblock \bibinfo{journal}{Energy Build.} \bibinfo{volume}{210},
  \bibinfo{pages}{109729}.
\bibitem[{Sonta and Jain(2020)}]{Sonta2020-pd}
\bibinfo{author}{Sonta, A.}, \bibinfo{author}{Jain, R.K.},
  \bibinfo{year}{2020}.
\newblock \bibinfo{title}{Learning socio-organizational network structure in
  buildings with ambient sensing data}.
\newblock \bibinfo{journal}{Data-Centric Engineering} \bibinfo{volume}{1},
  \bibinfo{pages}{e9}.
\bibitem[{Sood et~al.(2020)Sood, Janssen and Miller}]{Sood2020-vz}
\bibinfo{author}{Sood, T.}, \bibinfo{author}{Janssen, P.},
  \bibinfo{author}{Miller, C.}, \bibinfo{year}{2020}.
\newblock \bibinfo{title}{Spacematch: Using environmental preferences to match
  occupants to suitable activity-based workspaces}.
\newblock \bibinfo{journal}{Frontiers in Built Environment}
  \bibinfo{volume}{6}, \bibinfo{pages}{113}.
\bibitem[{Soomro et~al.(2021)Soomro, Bharathy, Biloria and
  Prasad}]{Soomro2021-ne}
\bibinfo{author}{Soomro, A.M.}, \bibinfo{author}{Bharathy, G.},
  \bibinfo{author}{Biloria, N.}, \bibinfo{author}{Prasad, M.},
  \bibinfo{year}{2021}.
\newblock \bibinfo{title}{A review on motivational nudges for enhancing
  building energy conservation behavior}.
\newblock \bibinfo{journal}{Journal of Smart Environments and Green Computing}
  \bibinfo{volume}{1}, \bibinfo{pages}{3--20}.
\bibitem[{Szczytowski(2014)}]{szczytowski2014geo}
\bibinfo{author}{Szczytowski, P.}, \bibinfo{year}{2014}.
\newblock \bibinfo{title}{Geo-fencing based disaster management service}, in:
  \bibinfo{booktitle}{Workshop on Agents, Virtual Societies and Analytics},
  \bibinfo{organization}{Springer}. pp. \bibinfo{pages}{11--21}.
\bibitem[{Tartarini et~al.(2023)Tartarini, Frei, Schiavon, Chua and
  Miller}]{Tartarini2023-iu}
\bibinfo{author}{Tartarini, F.}, \bibinfo{author}{Frei, M.},
  \bibinfo{author}{Schiavon, S.}, \bibinfo{author}{Chua, Y.X.},
  \bibinfo{author}{Miller, C.}, \bibinfo{year}{2023}.
\newblock \bibinfo{title}{Cozie apple: An {iOS} mobile and smartwatch
  application for environmental quality satisfaction and physiological data
  collection}.
\newblock \bibinfo{journal}{J. Phys. Conf. Ser.} \bibinfo{volume}{2600},
  \bibinfo{pages}{142003}.
\bibitem[{Tekler et~al.(2023)Tekler, Lei, Peng, Miller and
  Chong}]{Tekler2023-ri}
\bibinfo{author}{Tekler, Z.D.}, \bibinfo{author}{Lei, Y.},
  \bibinfo{author}{Peng, Y.}, \bibinfo{author}{Miller, C.},
  \bibinfo{author}{Chong, A.}, \bibinfo{year}{2023}.
\newblock \bibinfo{title}{A hybrid active learning framework for personal
  thermal comfort models}.
\newblock \bibinfo{journal}{Build. Environ.} \bibinfo{volume}{234},
  \bibinfo{pages}{110148}.
\bibitem[{Tobin et~al.(2023)Tobin, Heidari, Volpi, Sodder and
  Duncan}]{tobin2023use}
\bibinfo{author}{Tobin, K.}, \bibinfo{author}{Heidari, O.},
  \bibinfo{author}{Volpi, C.}, \bibinfo{author}{Sodder, S.},
  \bibinfo{author}{Duncan, D.}, \bibinfo{year}{2023}.
\newblock \bibinfo{title}{Use of geofencing interventions in population health
  research: a scoping review}.
\newblock \bibinfo{journal}{BMJ open} \bibinfo{volume}{13},
  \bibinfo{pages}{e069374}.
\bibitem[{Torresin et~al.(2024)Torresin, Al-Assaad, Aletta, Balderrama,
  Bivolarova, de~Souza, Dicle, Lee, Llorca-Bofí, Maula, Pigliautile, Pisello
  and Wu}]{Torresin2024PECS}
\bibinfo{author}{Torresin, S.}, \bibinfo{author}{Al-Assaad, D.},
  \bibinfo{author}{Aletta, F.}, \bibinfo{author}{Balderrama, A.},
  \bibinfo{author}{Bivolarova, M.P.}, \bibinfo{author}{de~Souza, L.P.},
  \bibinfo{author}{Dicle, S.Y.}, \bibinfo{author}{Lee, P.J.},
  \bibinfo{author}{Llorca-Bofí, J.}, \bibinfo{author}{Maula, H.},
  \bibinfo{author}{Pigliautile, I.}, \bibinfo{author}{Pisello, A.L.},
  \bibinfo{author}{Wu, Z.}, \bibinfo{year}{2024}.
\newblock \bibinfo{title}{Introducing the concept of acoustic personalised
  environmental control systems (acoustic pecs) within the framework of iea ebc
  annex 87}, in: \bibinfo{booktitle}{INTER-NOISE and NOISE-CON Congress and
  Conference Proceedings, INTER-NOISE24}, \bibinfo{publisher}{Institute of
  Noise Control Engineering}, \bibinfo{address}{Nantes, France}. pp.
  \bibinfo{pages}{3371--3375}.
\newblock \DOIprefix\doi{10.3397/IN_2024_3317}.
\bibitem[{Venema et~al.(2018)Venema, Kroese and De~Ridder}]{Venema2018-ft}
\bibinfo{author}{Venema, T.A.G.}, \bibinfo{author}{Kroese, F.M.},
  \bibinfo{author}{De~Ridder, D.T.D.}, \bibinfo{year}{2018}.
\newblock \bibinfo{title}{I'm still standing: A longitudinal study on the
  effect of a default nudge}.
\newblock \bibinfo{journal}{Psychology and Health} \bibinfo{volume}{33},
  \bibinfo{pages}{669--681}.
\bibitem[{Wang et~al.(2024)Wang, He, Zhai, Wang and Zhao}]{wang20243d}
\bibinfo{author}{Wang, Y.}, \bibinfo{author}{He, Z.}, \bibinfo{author}{Zhai,
  W.}, \bibinfo{author}{Wang, S.}, \bibinfo{author}{Zhao, C.},
  \bibinfo{year}{2024}.
\newblock \bibinfo{title}{How do the 3d urban morphological characteristics
  spatiotemporally affect the urban thermal environment? a case study of san
  antonio}.
\newblock \bibinfo{journal}{Building and Environment} \bibinfo{volume}{261},
  \bibinfo{pages}{111738}.
\bibitem[{Xu et~al.(2023)Xu, Yu, Sun and Tam}]{Xu2023-mi}
\bibinfo{author}{Xu, X.}, \bibinfo{author}{Yu, H.}, \bibinfo{author}{Sun, Q.},
  \bibinfo{author}{Tam, V.W.Y.}, \bibinfo{year}{2023}.
\newblock \bibinfo{title}{A critical review of occupant energy consumption
  behavior in buildings: How we got here, where we are, and where we are
  headed}.
\newblock \bibinfo{journal}{Renewable Sustainable Energy Rev.}
  \bibinfo{volume}{182}, \bibinfo{pages}{113396}.
\bibitem[{Yang et~al.(2021)Yang, Yang, Sun, Jin and Xiao}]{yang2021influence}
\bibinfo{author}{Yang, J.}, \bibinfo{author}{Yang, Y.}, \bibinfo{author}{Sun,
  D.}, \bibinfo{author}{Jin, C.}, \bibinfo{author}{Xiao, X.},
  \bibinfo{year}{2021}.
\newblock \bibinfo{title}{Influence of urban morphological characteristics on
  thermal environment}.
\newblock \bibinfo{journal}{Sustainable Cities and Society}
  \bibinfo{volume}{72}, \bibinfo{pages}{103045}.
\bibitem[{Yap et~al.(2023)Yap, Stouffs and Biljecki}]{2023_npjus_urbanity}
\bibinfo{author}{Yap, W.}, \bibinfo{author}{Stouffs, R.},
  \bibinfo{author}{Biljecki, F.}, \bibinfo{year}{2023}.
\newblock \bibinfo{title}{Urbanity: automated modelling and analysis of
  multidimensional networks in cities}.
\newblock \bibinfo{journal}{npj Urban Sustainability} \bibinfo{volume}{3}.
\newblock \DOIprefix\doi{10.1038/s42949-023-00125-w}.
\bibitem[{Zipf et~al.(2020)Zipf, Primack and Rothendler}]{Zipf2020-br}
\bibinfo{author}{Zipf, L.}, \bibinfo{author}{Primack, R.B.},
  \bibinfo{author}{Rothendler, M.}, \bibinfo{year}{2020}.
\newblock \bibinfo{title}{Citizen scientists and university students monitor
  noise pollution in cities and protected areas with smartphones}.
\newblock \bibinfo{journal}{PLoS One} \bibinfo{volume}{15},
  \bibinfo{pages}{e0236785}.

\end{thebibliography}
\end{document}